\documentclass[english,aip,jcp,preprint]{revtex4-1}
\usepackage[T1]{fontenc}
\usepackage[latin9]{inputenc}
\usepackage{array}
\usepackage{refstyle}
\usepackage{float}
\usepackage{mathrsfs}
\usepackage{amsmath}
\usepackage{graphicx}
\PassOptionsToPackage{version=3}{mhchem}
\usepackage{mhchem}

\makeatletter

\AtBeginDocument{\providecommand\figref[1]{\ref{fig:#1}}}
\AtBeginDocument{\providecommand\tabref[1]{\ref{tab:#1}}}
\AtBeginDocument{\providecommand\subsecref[1]{\ref{subsec:#1}}}
\newcommand{\lyxmathsym}[1]{\ifmmode\begingroup\def\b@ld{bold}
  \text{\ifx\math@version\b@ld\bfseries\fi#1}\endgroup\else#1\fi}

\providecommand{\tabularnewline}{\\}
\RS@ifundefined{subsecref}
  {\newref{subsec}{name = \RSsectxt}}
  {}
\RS@ifundefined{thmref}
  {\def\RSthmtxt{theorem~}\newref{thm}{name = \RSthmtxt}}
  {}
\RS@ifundefined{lemref}
  {\def\RSlemtxt{lemma~}\newref{lem}{name = \RSlemtxt}}
  {}

\usepackage{mhchem}
\usepackage[binary-units=true]{siunitx}
\usepackage[dvipsnames]{xcolor}
\usepackage{pgfplots}
\usepgfplotslibrary{groupplots}
\renewcommand{\tabref}{\Tabref}
\renewcommand{\figref}{\Figref}

\allowdisplaybreaks

\makeatother

\usepackage{babel}
\begin{document}
\title{A Critical Analysis of Least-Squares Tensor Hypercontraction Applied
to MP3}
\author{Devin A. Matthews}
\email{damatthews@smu.edu}

\affiliation{Southern Methodist University, Dallas, TX 75275, USA}
\begin{abstract}
The least-squares tensor hypercontraction (LS-THC) approach is a promising
method of reducing the high polynomial scaling of wavefunction methods,
for example those based on many-body perturbation theory or coupled
cluster. Here, we focus on LS-THC-MP3, and identify four variants
with differing error and efficiency characteristics. The performance
of LS-THC-MP3 is analyzed for regular test systems with up to 40 first-row
atoms. We also analyze the size-extensivity/size-consistency and grid-
and basis set-dependence of LS-THC-MP3. Overall, the errors observed
are favorably small in comparison with standard density fitting, and
a more streamlined method of generating grids via pruning is suggested.
Practical crossover (the point at which LS-THC-MP3 is cheaper than
the canonical method) is achieved around 240 correlated electrons.
Despite several drawbacks of LS-THC that have been identified: a small
but non-zero size-consistency error, poor description of angular correlation,
and potentially large increase of error with basis set size, the results
show that LS-THC has significant potential for practical application
to MP3 and other wavefunction methods.
\end{abstract}
\maketitle

\section{Introduction}

The problem of the steep polynomial scaling of the computational cost
of quantum chemical methods, in particular those based on explicit
wavefunction representations, is one of the major roadblocks in extending
accurate computational methods to large and complex systems. It has
long been known that electronic interactions (in neutral systems),
including dynamical electron correlation, are inherently ``near-sighted''.\citep{kohnDensityFunctionalDensity1996}
This fact has led to the development of a number of localized electron
correlation methods, including those based on pair natural orbitals,\citep{hattigLocalExplicitlyCorrelated2012,riplingerEfficientLinearScaling2013,schwilkScalableElectronCorrelation2017,guoCommunicationImprovedLinear2018}
molecular fragmentation,\citep{kitauraFragmentMolecularOrbital1999,gordonEffectiveFragmentPotential2001}
Hilbert-space localization,\citep{rolikGeneralorderLocalCoupledcluster2011,rolikEfficientLinearscalingCCSD2013,eriksenLinearScalingCoupledCluster2015,liClusterinmoleculeLocalCorrelation2016}
atomic orbital representations,\citep{ochsenfeldLinearScalingMethodsQuantum2007}
and others. For total energies, such approaches have been very successful,
end such techniques have even been extended to excited state energies
and other properties,\citep{duttaPairNaturalOrbital2016,duttaExploringAccuracyLow2018,pengStateAveragedPairNatural2018,pinskiCommunicationExactAnalytical2018}
albeit with some limitations.

In another direction, there has been less vigorous research into the
fact that electronic interactions are inherently structured. In this
viewpoint, one considers a global representation of electronic interactions,
either the two-electron Hamiltonian or the parameters of the wavefunction
ansatz (e.g. cluster amplitudes), and attempts to discover approximate
forms which reduce the dimensionality and scaling of the correlation
problem. A simple example is given by the CANDECOMP/PARAFAC\citep{koldaTensorDecompositionsApplications2009}
(CP---later backronymed to canonical polyadic\citep{kiersStandardizedNotationTerminology2000})
decomposition applied to the two-electron part of the Hamiltonian,
$\hat{G}$, expressed as a 4-dimensional tensor,
\[
g_{\rho\sigma}^{\mu\nu}\equiv\langle\mu\nu|\hat{g}|\rho\sigma\rangle\approx\sum_{r=1}^{R}U_{\mu r}V_{\nu r}W_{\rho r}X_{\sigma r}
\]
where permutational symmetry has not been imposed. Note that the non-antisymmetrized
integrals are used here. Instead of a single 4-dimensional tensor,
one only has to deal with four matrices, and assuming that the number
of factors, $R$, grows linearly with molecular size, reduced scaling
both in storage and in computation. While the CP decomposition provides
a relatively simple example of tensor decompositions, and how they
can lead to reduction in computational scaling,\citep{benediktTensorDecompositionPostHartree2011}
a practical and feasible implementation of CP for large molecules
has not as yet been developed. 

A somewhat more promising alternative is the tensor hypercontraction
(THC) approach,\citep{hohensteinTensorHypercontractionDensity2012}
especially the least-squares (LS-THC) variant.\citep{parrishTensorHypercontractionII2012}
The THC factorization is itself similar to but slightly more complex
than the CP format, having now five matrix components, but the real
advantage of LS-THC is that four out of these five matrices can be
trivial computed given a suitable set of real-space grid points. The
remaining matrix can be determined in closed form by fitting the Hamiltonian
(or wavefunction components). In contrast, determining the CP or direct
THC decomposition is a highly non-linear process which is fraught
with convergence problems.\citep{koldaTensorDecompositionsApplications2009,schutskiTensorstructuredCoupledCluster2017,hummelLowRankFactorization2017}
LS-THC, applied the the Hamiltonian, is exceptionally effective in
accurately reproducing e.g. MP2 energies with rather small grids\citep{kokkilaschumacherTensorHypercontractionSecondOrder2015}
(scaling linearly with system size, and with a number of points similar
to or fewer than typical auxiliary basis sets used for density fitting).
Largely, we can attribute this success to the fact that the bare Coulomb
interaction is entirely local (in a different sense than for local
correlation methods): an integral over the Coulomb operator can be
exactly computed using a real-space quadrature over a grid. This same
property is responsible for the favorable computational properties
of local density functional methods as well as density fitting/resolution-of-the-identity
approaches.

The wavefunction, however, is another story. Taking the doubles amplitudes
from configuration interaction (CI) or, neglecting for a moment higher-order
terms, from coupled cluster (CC), we can consider these as integrals
over a ``two-electron interaction kernel'',
\[
c_{ij}^{ab}\sim t_{ij}^{ab}=\langle ab|\hat{\tau}|ij\rangle
\]
At first order in Møller-Plesset perturbation theory, and we have,
\[
(t^{[1]})_{ij}^{ab}=\frac{g_{ij}^{ab}}{\epsilon_{i}+\epsilon_{j}-\epsilon_{a}-\epsilon_{b}}
\]
Numerical experiments confirm that these amplitudes have both linear
(or at least sub-quadratic) CP rank\citep{benediktTensorDecompositionPostHartree2013}
as well as linear SVD rank (formatted as an $ai\times bj$ matrix).\citep{parrishRankReducedCoupled2019}
Initial applications of LS-THC to cluster amplitudes also support
a low-rank structure,\citep{hohensteinCommunicationTensorHypercontraction2012,schutskiTensorstructuredCoupledCluster2017}
which are confirmed by results in this work. The CP rank of the orbital
energy denominators is actually constant, as evidenced by the Laplace
transform technique.\citep{katsUseLaplaceTransform2008} In total,
the cluster amplitudes appear to be manifestly factorizable.

Nonetheless, an important difference remains: the two-electron interaction
kernel is not a local operator. In this work, we demonstrate numerically
that the LS-THC decomposition is inherently less accurate for double
amplitudes, compared to Coulomb integrals. Additionally, we show that
the form of LS-THC necessary for treating iterative methods like CCSD
incurs additional errors due to forced approximation of exchange integrals.
This is in sharp contrast to the expectations of Hohenstein et al.
that, ``If more fidelity is required in the correlation energies,
it is possible to simply use denser grids.''\citep{hohensteinCommunicationTensorHypercontraction2012}
While this statement seems reasonable given that the LS-THC decomposition
does indeed become exact when $N(N+1)/2$ grid points are employed,
we show that errors for practical grid sizes may be large and slowly
decreasing. Despite this theoretical shortcoming of LS-THC, we do
show that errors in MP3 energies when employing LS-THC-decomposed
$\hat{T}_{2}^{[1]}$ or $\hat{T}_{2}^{[2]}$ amplitudes can be made
quite reasonable. Most importantly, we show that our implementation
of LS-THC-MP3 achieves practical crossover around 240 correlated electrons,
a point which will steadily move towards smaller systems as grid size
and implementation efficiency are improved.

\section{Theory}

The LS-THC decomposition imposes a CP-like structure on the two-electron
Hamiltonian elements,\citep{hohensteinTensorHypercontractionDensity2012}
\[
g_{\rho\sigma}^{\mu\nu}\approx\sum_{RS=1}^{n_{P}}X_{\mu}^{R}X_{\rho}^{R}V_{RS}X_{\nu}^{S}X_{\sigma}^{S}
\]
where $n_{P}$ is the number of grid-points. We have used the standard
notation for atomic orbital (AO) and molecular orbital (MO) indices,
with $\mu\nu\rho\sigma$ denoting AOs, $abcdef$ denoting virtual
MOs, and $ijklmn$ denoting occupied MOs. Capital Roman letters $RSTUVWXY$
will be used to denote grid points; it should be noted that we will
also make use of $\mathbf{T}$, $\mathbf{V}$, and $\mathbf{X}$ matrices,
with elements such as $T_{RS}$, $V_{UW}$, $X_{\mu}^{Y}$ etc. (as
above), and that these should not be confused with grid indices.

In this work, we use grid points pruned from an initial Becke-style\citep{beckeMulticenterNumericalIntegration1988}
molecular grid using the Cholesky decomposition approach of Ref.~\citenum{matthewsImprovedGridOptimization2020}.
The collocation matrix $\mathbf{X}$ is determined by simple evaluation
of the orbitals at the grid points, weighted by the fourth root of
the grid quadrature weights. The LS-THC decomposition also applies
directly to integrals in the molecular orbital basis,
\[
g_{rs}^{pq}\approx\sum_{R=1}^{n_{pr}}\sum_{S=1}^{n_{qs}}(X^{(pr)})_{p}^{R}(X^{(pr)})_{r}^{R}V_{RS}^{(pqrs)}(X^{(qs)})_{q}^{S}(X^{(qs)})_{s}^{S}
\]
where now we may have one of three grids determining $\{\mathbf{X}^{(ab)},\mathbf{X}^{(ai)},\mathbf{X}^{(ij)}\}$
collocation matrices with $\{n_{ab},n_{ai},n_{ij}\}$ grid points,
respectively. $\mathbf{V}$ is determined uniquely for each of the
$abcd$, $abci$, $abij$, $aibj$, $aijk$, and $ijkl$ MO distributions
(note these are in Dirac order). These indices on $\mathbf{X}$, $\mathbf{V}$,
and $n$ only convey the relevant MO distributions (occupied/virtual)
and are not indices in the usual sense. In each case, the ``core''
matrix $\mathbf{V}$ is determined by a least-squares fit,\citep{parrishTensorHypercontractionII2012}
\[
V_{RS}=\sum_{TU=1}^{n_{P}}S_{RT}^{-1}X_{\mu}^{T}X_{\rho}^{T}g_{\rho\sigma}^{\mu\nu}X_{\nu}^{U}X_{\sigma}^{U}S_{US}^{-1}
\]
with $S_{RS}=\sum_{\mu\nu}X_{\mu}^{R}X_{\nu}^{R}X_{\mu}^{S}X_{\nu}^{S}$,
and $S_{RS}^{-1}$ is an element of the matrix inverse of $\mathbf{S}$.
We have used the AO integrals here for example; the procedure is the
same for the MO integrals. Note that having three MO $\mathbf{X}$
matrices also implies three distinct $\mathbf{S}$ matrices, with
similar notation. Of course each MO distribution may (and should)
be fit independently. In this work, we employ a pivoted low-rank Cholesky
decomposition of $\mathbf{S}$, and the explicit inverse is replaced
by two triangular solves (on each side). It is important that the
$\mathbf{V}$ matrix obtained this way be explicitly symmetrized,
as the values of ``less important'' elements are poorly determined
numerically. Determining $\mathbf{V}$ from the exact AO or MO integrals
scales as $\mathscr{\mathcal{O}}(N^{5})$, where $N$ is a measure
of the system size, not considering sparsity. Using density-fitted
integrals\citep{aquilanteUnbiasedAuxiliaryBasis2007} (or RI/CD integrals\citep{kendallImpactResolutionIdentity1997,kochReducedScalingElectronic2003,aquilanteLowcostEvaluationExchange2007})
reduces this to $\mathscr{\mathcal{O}}(N^{4})$.

In order to employ the LS-THC factorization in MP2, an additional
approximation must be made to eliminate the inseparability of the
orbital energy denominators. This is commonly handled by the Laplace
transform approach,\citep{katsUseLaplaceTransform2008} where a numerical
quadrature is used,
\begin{align*}
x^{-1}=\int_{0}^{\infty}e^{-xt}dt & \approx\sum_{l=1}^{L}w_{l}e^{-xt_{l}}\quad(x>0)\\
-(\epsilon_{a}+\epsilon_{b}-\epsilon_{i}-\epsilon_{j})^{-1} & \approx-\sum_{l=1}^{L}g_{a}^{l}g_{b}^{l}g_{i}^{l}g_{j}^{l}
\end{align*}
where $w_{l}$ and $t_{l}$ are the weights and abscissas of the quadrature,
and $g_{a}^{l}=w_{l}^{1/4}e^{-\epsilon_{a}t_{l}}$, $g_{i}^{l}=w_{l}^{1/4}e^{\epsilon_{i}t_{l}}$.
Only a small, constant number of quadrature points are required; in
this work we use 9 points (cc-pVDZ) or 10 points (cc-pVTZ). The error
introduced by the Laplace transform with this number of points is
essentially negligible.

Now, the LS-THC-MP2 energy expression can be constructed,
\begin{align*}
E_{LS-THC-MP2} & =-2\sum_{abij}\sum_{RSTU=1}^{n_{ai}}\sum_{l=1}^{L}X_{a}^{R}X_{i}^{R}V_{RS}X_{b}^{S}X_{j}^{S}g_{a}^{l}g_{b}^{l}g_{i}^{l}g_{j}^{l}X_{a}^{T}X_{i}^{T}V_{TU}X_{b}^{U}X_{j}^{U}\\
 & \quad+\sum_{abij}\sum_{RSTU=1}^{n_{ai}}\sum_{l=1}^{L}X_{a}^{R}X_{i}^{R}V_{RS}X_{b}^{S}X_{j}^{S}g_{a}^{l}g_{b}^{l}g_{i}^{l}g_{j}^{l}X_{a}^{T}X_{j}^{T}V_{TU}X_{b}^{U}X_{i}^{U}\\
 & =E_{C}+E_{X}
\end{align*}
where $\mathbf{X}$ and $\mathbf{V}$ are both those specific to the
$ai$ and $abij$ MO distributions only, and the bounds on the MO
summations are implicit. We have not shown the full factorization
here; the first, Coulomb, term can be computed in $\mathcal{O}(N^{3})$
time and the second, eXchange, term in $\mathcal{O}(N^{4})$ time.
Note that nearly all terms must be recomputed for each Laplace quadrature
point, so that using fewer points will drastically reduce the running
time.

\begin{figure}
\begin{centering}
\includegraphics[width=0.5\textwidth]{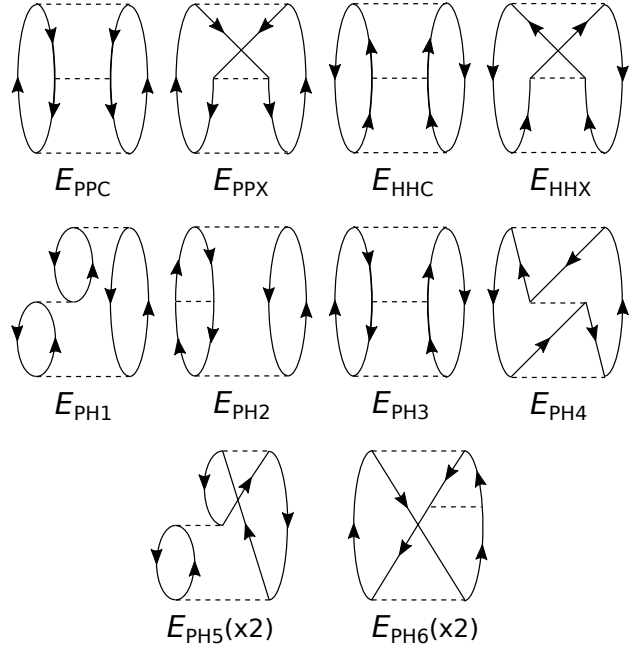}
\par\end{centering}
\caption{\label{fig:Goldstone-diagrams-for}Goldstone diagrams for the MP3
energy and their associated energy contributions. The last two energy
contributions include two diagrams each, related by a vertical reflection.
Denominator lines are not shown for clarity.}
\end{figure}

The LS-THC-MP3 energy expression is significantly more involved, as
there are 10 unique terms, depicted diagrammatically in \figref{Goldstone-diagrams-for}.
As an example, consider the $E_{\text{PPC}}$ term,
\begin{align*}
E_{\text{PPC}} & =2\sum_{abijef}\sum_{RSTU=1}^{n_{ai}}\sum_{VW=1}^{n_{ab}}\sum_{lm=1}^{L}(X^{(ai)})_{a}^{R}(X^{(ai)})_{i}^{R}V_{RS}^{(abij)}(X^{(ai)})_{b}^{S}(X^{(ai)})_{j}^{S}\\
 & \quad\quad\quad\quad\quad\quad\quad\times g_{a}^{l}g_{b}^{l}g_{i}^{l}g_{j}^{l}(X^{(ab)})_{a}^{V}(X^{(ab)})_{e}^{V}V_{VW}^{(abcd)}(X^{(ab)})_{b}^{W}(X^{(ab)})_{f}^{W}\\
 & \quad\quad\quad\quad\quad\quad\quad\times g_{e}^{m}g_{f}^{m}g_{i}^{m}g_{j}^{m}(X^{(ai)})_{e}^{T}(X^{(ai)})_{i}^{T}V_{TU}^{(abij)}(X^{(ai)})_{f}^{U}(X^{(ai)})_{j}^{U}
\end{align*}
Unlike in LS-THC-MP2, we must now deal with multiple distinct $\mathbf{X}$
and $\mathbf{V}$ matrices. The factorization of these equations can
also be accomplished such that $\mathcal{O}(N^{4})$ scaling is achieved,
and multiple such factorizations have been published in the literature.
We use a different formulation that aims to minimize the number of
$\mathcal{O}(N^{4})$ steps while also minimizing memory usage and
creating opportunities for index blocking and loop fusion. We have
not included the full factored equations here as work on improved
factorization is ongoing and the working equations are rapidly evolving.
One of the largest obstacles to a practical implementation of LS-THC-MP3
is the introduction of two Laplace transform quadratures, which introduces
an $L^{2}$ dependence into the computational cost. This cost multiplier
ranges from $\sim\!10\times$ (minimal number of points) to $\sim\!100\times$
(this work).

Instead of including an explicit dependence on the orbital energy
denominators in the MP3 energy, one can instead substitute the first-order
cluster amplitudes, $\hat{T}_{2}^{[1]}$, and perform an LS-THC decomposition
as for the two-electron integrals. Doing so completely removes the
Laplace transform factors, but otherwise leaves the structure of the
equations unchanged, e.g.,
\begin{align*}
E_{\text{PPC}} & =2\sum_{abijef}\sum_{RSTU=1}^{n_{ai}}\sum_{VW=1}^{n_{ab}}(X^{(ai)})_{a}^{R}(X^{(ai)})_{i}^{R}T_{RS}^{[1]}(X^{(ai)})_{b}^{S}(X^{(ai)})_{j}^{S}\\
 & \quad\quad\quad\quad\quad\quad\quad\quad\;\times(X^{(ab)})_{a}^{V}(X^{(ab)})_{e}^{V}V_{VW}^{(abcd)}(X^{(ab)})_{b}^{W}(X^{(ab)})_{f}^{W}\\
 & \quad\quad\quad\quad\quad\quad\quad\quad\;\times(X^{(ai)})_{e}^{T}(X^{(ai)})_{i}^{T}T_{TU}^{[1]}(X^{(ai)})_{f}^{U}(X^{(ai)})_{j}^{U}
\end{align*}
where the LS-THC decomposition of the amplitudes is given by,
\[
(t^{[1]})_{ij}^{ab}\approx\sum_{RS=1}^{n_{ai}}(X^{(ai)})_{a}^{R}(X^{(ai)})_{i}^{R}T_{RS}^{[1]}(X^{(ai)})_{b}^{S}(X^{(ai)})_{j}^{S}
\]
This simplification of the LS-THC-MP3 equations was simultaneously
recognized by Lee et al.\citep{leeSystematicallyImprovableTensor2020}
Because this new method introduces an additional approximation of
the cluster amplitudes (which were implicit and ``exact'' in the
original formulation\citep{hohensteinTensorHypercontractionDensity2012}),
the energy obtained will not be the same. We denote these two methods
as LS-THC-MP3a and LS-THC-MP3b, respectively. A similar substitution
can also be made for the LS-THC-MP2 energy, which reduces the cost
(of $\mathcal{O}(N^{4})$ steps) by a factor of $L$. We then denote
LS-THC-MP2a and LS-THC-MP2b methods in a similar fashion. When there
is little possibility of confusion, we will refer to these LS-THC
methods simply as MP2a, MP3b, etc.

MP3, by itself, is not a particularly useful method. It suffers from
intruder states and does not include the effect of higher-order excitations
through the singles and doubles (e.g. from SDQ-MP4) or the infinite-order
correlation effects of CI or CC. But, it is an important stepping
stone to CC methods such as CCSD, as it captures the primary (linear)
terms of the CCSD iteration. In fact, LCCD (also known as CEPA0\citep{kochComparisonCEPACPMET1980})
is simply an iterated version of MP3. The main difference is in the
energy expression: while the full MP3 energy expression can be completely
expanded as in MP3a, the LCCD (or general CC) energy must have the
form of MP2b, where the first-order amplitudes are replaced by the
converged ones. In each LCCD iteration, we must also directly solve
for the updated cluster amplitudes. Thus, decomposition of the amplitudes
is absolutely mandatory in moving to iterative methods.

\begin{figure}
\begin{centering}
\includegraphics[width=0.5\textwidth]{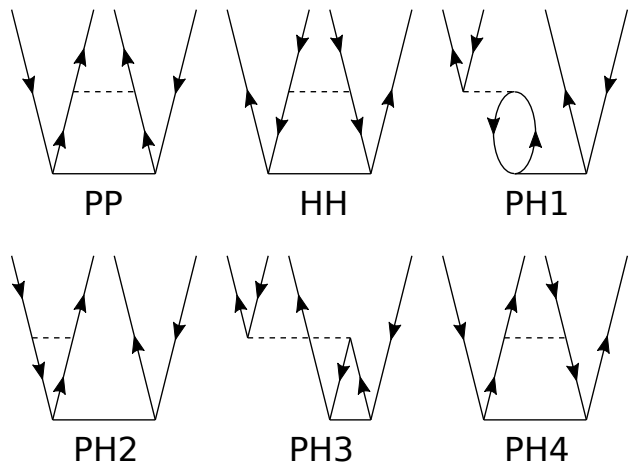}
\par\end{centering}
\caption{\label{fig:Goldstone-diagrams-for-1}Open Goldstone diagrams for an
LCCD iteration, or equivalently for determining the second-order amplitudes
in MP3 (denominator lines not shown). Each term is labeled for ease
of reference.}
\end{figure}

In order to approximate this situation in MP3, we can reformulate
the MP3 energy as a single iteration of LCCD: the starting amplitudes
are the first-order ones, and the final amplitudes from which the
energy are computed are simply the second-order ones. Each LCCD iteration
comprises six open Goldstone diagrams, depicted in \figref{Goldstone-diagrams-for-1}.
Note that we assume a canonical Hartree--Fock reference, and that
the diagonal elements of the Fock matrix are included via the orbital
energy denominators. These are open diagrams that, with division by
the orbital energy differences, determine the updated (for MP3, second-order)
amplitudes. Since we do not want to explicitly form the full amplitude
tensor, we may insert a Laplace transform of the denominators and
fit the result to an LS-THC form as above. Using the PP term as an
example,
\begin{align*}
T_{RS}^{[2]} & \gets-\sum_{abefij}\sum_{TUVW=1}^{n_{ai}}\sum_{XY=1}^{n_{ab}}\sum_{l=1}^{L}S_{RT}^{-1}(X^{(ai)})_{a}^{T}(X^{(ai)})_{i}^{T}(X^{(ai)})_{b}^{U}(X^{(ai)})_{j}^{U}g_{a}^{l}g_{b}^{l}g_{i}^{l}g_{j}^{l}\\
 & \quad\quad\quad\quad\quad\quad\quad\quad\quad\quad\quad\quad\;\;\times(X^{(ab)})_{a}^{X}(X^{(ab)})_{e}^{X}V_{XY}^{(abcd)}(X^{(ab)})_{b}^{Y}(X^{(ab)})_{f}^{Y}\\
 & \quad\quad\quad\quad\quad\quad\quad\quad\quad\quad\quad\quad\;\;\times(X^{(ai)})_{e}^{V}(X^{(ai)})_{i}^{V}T_{VW}^{[1]}(X^{(ai)})_{f}^{W}(X^{(ai)})_{j}^{W}S_{US}^{-1}
\end{align*}
Each of these terms may also be factorized such that the scaling is
at most $\mathcal{O}(N^{4})$. Substituting $\mathbf{T}^{[2]}$ for
$\mathbf{T}^{[1]}$ in the MP2b energy expression then gives a new
variant of MP3 which we term MP3c.

Of course, the determination of the $\mathbf{T}^{[2]}$ matrix now
has a dependence on $L$, which increases the cost considerably compared
to MP3b. Instead, we may first fit the LCCD residual vector, and then
perform the orbital energy weighting separately (again using the PP
term as an example),
\begin{align*}
Z_{RS} & \gets\sum_{abefij}\sum_{TUVW=1}^{n_{ai}}\sum_{XY=1}^{n_{ab}}S_{RT}^{-1}(X^{(ai)})_{a}^{T}(X^{(ai)})_{i}^{T}(X^{(ai)})_{b}^{U}(X^{(ai)})_{j}^{U}\\
 & \quad\quad\quad\quad\quad\quad\quad\quad\quad\quad\times(X^{(ab)})_{a}^{X}(X^{(ab)})_{e}^{X}V_{XY}^{(abcd)}(X^{(ab)})_{b}^{Y}(X^{(ab)})_{f}^{Y}\\
 & \quad\quad\quad\quad\quad\quad\quad\quad\quad\quad\times(X^{(ai)})_{e}^{V}(X^{(ai)})_{i}^{V}T_{VW}^{[1]}(X^{(ai)})_{f}^{W}(X^{(ai)})_{j}^{W}S_{US}^{-1}\\
T_{RS}^{[2]} & =-\sum_{abij}\sum_{TUVW=1}^{n_{ai}}\sum_{l=1}^{L}S_{RT}^{-1}(X^{(ai)})_{a}^{T}(X^{(ai)})_{i}^{T}(X^{(ai)})_{b}^{U}(X^{(ai)})_{j}^{U}g_{a}^{l}g_{b}^{l}g_{i}^{l}g_{j}^{l}\\
 & \quad\quad\quad\quad\quad\quad\quad\quad\quad\;\;\times(X^{(ai)})_{a}^{V}(X^{(ai)})_{i}^{V}Z_{VW}(X^{(ai)})_{b}^{W}(X^{(ai)})_{j}^{W}S_{US}^{-1}
\end{align*}
The second step only scales as $\mathcal{O}(N^{3})$, and now the
first step, which dominates the total cost, no longer depends on $L$.
We denote this final, ``double-fitted'' variant as MP3d. The orbital
weighting step is precisely the same as that for determining $\mathbf{T}^{[1]}$,
except that $\mathbf{Z}$ is inserted instead of $\mathbf{V}^{(abij)}$.

MP2b and MP3d are alike in that they both formulate the energy as
a contraction between an approximate amplitude tensor and the Coulomb
and exchange integrals,
\begin{align*}
E_{n+1} & =-2\sum_{abij}\sum_{RSTU=1}^{n_{ai}}X_{a}^{R}X_{i}^{R}T_{RS}^{[n]}X_{b}^{S}X_{j}^{S}X_{a}^{T}X_{i}^{T}V_{TU}X_{b}^{U}X_{j}^{U}\\
 & \quad+\sum_{abij}\sum_{RSTU=1}^{n_{ai}}X_{a}^{R}X_{i}^{R}T_{RS}^{[n]}X_{b}^{S}X_{j}^{S}X_{a}^{T}X_{j}^{T}V_{TU}X_{b}^{U}X_{i}^{U}
\end{align*}
where $n$ is either 1 (LS-THC-MP2b) or 2 (LS-THC-MP3d), and the $\mathbf{X}^{(ai)}$
grid is assumed. Furthermore, we can explicitly write the $\mathbf{T}^{[n]}$
core matrix in each case by LS-THC fitting of either the integrals
or the LCCD residual, respectively,
\begin{align*}
E_{n+1} & =-2\sum_{abij}\sum_{RSTU=1}^{n_{ai}}\left(\sum_{efmn}\sum_{VWXY=1}^{n_{ai}}\sum_{l=1}^{L}X_{e}^{V}X_{m}^{V}Z_{VW}^{[n]}X_{f}^{W}X_{n}^{W}g_{e}^{l}g_{f}^{l}g_{m}^{l}g_{n}^{l}\right.\\
 & \quad\quad\times\left.X_{e}^{X}X_{m}^{X}X_{f}^{Y}X_{n}^{Y}S_{XR}^{-1}S_{YS}^{-1}\vphantom{\sum_{VWXY=1}^{n_{ai}}}\right)X_{a}^{R}X_{i}^{R}X_{b}^{S}X_{j}^{S}X_{a}^{T}X_{i}^{T}V_{TU}X_{b}^{U}X_{j}^{U}\\
 & \quad+\sum_{abij}\sum_{RSTU=1}^{n_{ai}}\left(\sum_{efmn}\sum_{VWXY=1}^{n_{ai}}\sum_{l=1}^{L}X_{e}^{V}X_{m}^{V}Z_{VW}^{[n]}X_{f}^{W}X_{n}^{W}g_{e}^{l}g_{f}^{l}g_{m}^{l}g_{n}^{l}\right.\\
 & \quad\quad\times\left.X_{e}^{X}X_{m}^{X}X_{f}^{Y}X_{n}^{Y}S_{XR}^{-1}S_{YS}^{-1}\vphantom{\sum_{VWXY=1}^{n_{ai}}}\right)X_{a}^{R}X_{i}^{R}X_{b}^{S}X_{j}^{S}X_{a}^{T}X_{j}^{T}V_{TU}X_{b}^{U}X_{i}^{U}
\end{align*}
where the parentheses group $\mathbf{T}^{[n]}$, and $\mathbf{Z}^{[n]}$
is either $\mathbf{V}$ ($n=1$) or $\mathbf{Z}$ ($n=2$). From this
form, we can regroup quantities and arrive at a different, but numerically
equivalent interpretation,
\begin{align*}
E_{n+1} & =-2\sum_{efmn}\sum_{VWXY=1}^{n_{ai}}\sum_{l=1}^{L}X_{e}^{V}X_{m}^{V}Z_{VW}^{[n]}X_{f}^{W}X_{n}^{W}g_{e}^{l}g_{f}^{l}g_{m}^{l}g_{n}^{l}X_{e}^{X}X_{m}^{X}X_{f}^{Y}X_{n}^{Y}\\
 & \quad\quad\times\left(\sum_{abij}\sum_{RSTU=1}^{n_{ai}}S_{XR}^{-1}S_{YS}^{-1}X_{a}^{R}X_{i}^{R}X_{b}^{S}X_{j}^{S}X_{a}^{T}X_{i}^{T}V_{TU}X_{b}^{U}X_{j}^{U}\right)\\
 & \quad+\sum_{efmn}\sum_{VWXY=1}^{n_{ai}}\sum_{l=1}^{L}X_{e}^{V}X_{m}^{V}Z_{VW}^{[n]}X_{f}^{W}X_{n}^{W}g_{e}^{l}g_{f}^{l}g_{m}^{l}g_{n}^{l}X_{e}^{X}X_{m}^{X}X_{f}^{Y}X_{n}^{Y}\\
 & \quad\quad\times\left(\sum_{abij}\sum_{RSTU=1}^{n_{ai}}S_{XR}^{-1}S_{YS}^{-1}X_{a}^{R}X_{i}^{R}X_{b}^{S}X_{j}^{S}X_{a}^{T}X_{j}^{T}V_{TU}X_{b}^{U}X_{i}^{U}\right)\\
 & =-2\sum_{efmn}\sum_{VWXY=1}^{n_{ai}}\sum_{l=1}^{L}X_{e}^{V}X_{m}^{V}Z_{VW}^{[n]}X_{f}^{W}X_{n}^{W}g_{e}^{l}g_{f}^{l}g_{m}^{l}g_{n}^{l}X_{e}^{X}X_{m}^{X}V_{XY}X_{f}^{Y}X_{n}^{Y}\\
 & \quad+\sum_{efmn}\sum_{VWXY=1}^{n_{ai}}\sum_{l=1}^{L}X_{e}^{V}X_{m}^{V}Z_{VW}^{[n]}X_{f}^{W}X_{n}^{W}g_{e}^{l}g_{f}^{l}g_{m}^{l}g_{n}^{l}X_{e}^{X}X_{m}^{X}\tilde{V}_{XY}X_{f}^{Y}X_{n}^{Y}
\end{align*}
In the first term, refitting $\mathbf{V}$ is an idempotent operation.
In fact, the Coulomb term for MP2b is precisely equal to that in MP2a
(except for the introduction of additional accumulated round-off error).
However, in the second term, we have now performed a refitting of
$\mathbf{V}$, but in the exchange ordering. Thus, we are attempting
to fit a purely non-local operator with an LS-THC decomposition. Even
if $\mathbf{Z}$ were a purely local operator (as in MP2b), we would
still experience error due to this exchange fitting. Because all iterative
CI and CC methods have an energy expression of this form, they will
all necessarily experience an ``exchange fitting error'', although
numerical results below show that the effect is not necessarily uniform.

We should make one final note about the grids used. The Cholesky decomposition-based
pruning procedure in Ref.~\citenum{matthewsImprovedGridOptimization2020}
naturally produces grids of different sizes for each of the three
MO pair distributions from the same parent grid. The size of the final
grids is controlled by varying a cutoff parameter $\epsilon$, or
conversely by fixing $\epsilon$ and varying the parent grid size.
We use the latter approach here (except for the results in \figref{grids2}),
with a fixed $\epsilon=10^{-5}$. The size of the parent grid is specified
by the integer triplet $(L_{max},N_{1},N_{H})$, where $L_{max}$
is the highest angular momentum for which any of the angular grids
is an exact quadrature, $N_{1}$ specifies the number of radial quadrature
points for first-row atoms, and $N_{H}$ specifies the number of radial
quadrature points for hydrogen atoms. All other aspects of constructing
the grids are detailed in Ref.~\citenum{matthewsImprovedGridOptimization2020}.

\section{Results and Discussion}

In order to understand the intrinsic errors associated with each LS-THC-MP2
and LS-THC-MP3 variant, we performed a series of calculations on regular
test systems of increasing size: water clusters, linear alkanes, and
linear alkenes, each with up to 40 first-row atoms. Density fitting
was used for the reference calculations as well as the LS-THC calculations
(so that, technically, these are LS-DF-THC calculations). Previous
work has shown that the density fitting and LS-THC errors are essentially
additive,\citep{parrishTensorHypercontractionII2012} so we consider
the $E_{\text{LS-THC-DF}}-E_{\text{DF}}$ difference to be representative
of the true LS-THC error. MP3a, MP3c, and canonical MP3 (without density
fitting) calculations were only completed through systems with 20
first-row atoms. MP3 errors are for the MP3 contribution to the correlation
energy only (i.e. excluding the MP2 energy). The cc-pVDZ basis set\citep{dunningGaussianBasisSets1989}
was used for all calculations except where indicated, and the frozen
core approximation was used throughout. Corresponding cc-pV$X$Z-RI
auxiliary basis sets\citep{weigendEfficientUseCorrelation2002} were
used for the correlation energy. Unless otherwise noted, a (7,19,11)
parent grid was used with $\epsilon=10^{-5}$ (for comparison, the
common SG-1 grid\citep{gillStandardGridDensity1993} is roughly similar
to the much larger (23,51,43) grid as used in this work). In all cases
point group symmetry is not used, except for determining the orientation
of the molecule. All calculations (including DF-MP2 and DF-MP3) were
performed with a development version of CFOUR.\citep{matthewsCoupledclusterTechniquesComputational2020}
We initially tried to use other packages for the density-fitting calculations,
but were unable to compete calculations on the larger systems due
to limitations of the 32-bit BLAS interface used.

\subsection{\label{subsec:Water-Clusters}Water Clusters}

\begin{figure}
\begin{tikzpicture}
\pgfplotsset{every axis plot/.append style={very thick}}
\begin{axis}[
scale=1.5,
xmin=1, xmax=40,
ymin=-2, ymax=2,
xlabel={$n$},
ylabel={Error (kcal/mol)},
legend style={at={(1.05,0.5)},anchor=west,legend cell align=right}
]
\addplot[draw=blue] table[x=n,y=DF-MP2] {h2o-err.dat};
\addlegendentry{DF-MP2}
\addplot[draw=blue,dotted,domain=20:40,forget plot] {0.0265*x-0.0129};
\addplot[draw=black!50!white] table[x=n,y=MP2a] {h2o-err.dat};
\addlegendentry{LS-THC-MP2a}
\addplot[draw=yellow] table[x=n,y=MP3a] {h2o-err.dat};
\addlegendentry{LS-THC-MP3a}
\addplot[draw=yellow,dotted,domain=20:40,forget plot] {0.0002*x-0.0003};
\addplot[draw=cyan] table[x=n,y=MP3b] {h2o-err.dat};
\addlegendentry{LS-THC-MP3b}
\addplot[draw=red] table[x=n,y=MP2b] {h2o-err.dat};
\addlegendentry{LS-THC-MP2b}
\addplot[draw=green!50!black] table[x=n,y=MP3c] {h2o-err.dat};
\addlegendentry{LS-THC-MP3c}
\addplot[draw=green!50!black,dotted,domain=20:40,forget plot] {-0.042*x+0.4087};
\addplot[draw=purple] table[x=n,y=MP3d] {h2o-err.dat};
\addlegendentry{LS-THC-MP3d}
\addplot[draw=orange] table[x=n,y=DF-MP3] {h2o-err.dat};
\addlegendentry{DF-MP3}
\addplot[draw=orange,dotted,domain=20:40,forget plot] {-0.0683*x+0.0014};
\end{axis}
\end{tikzpicture}

\caption{\label{fig:Errors-for-water}Errors for water clusters, $\ce{(H2O)}_{n}$.
Note that legend entries are given in the same order as the error
values at $n=20$, and that the curves for LS-THC-MP2a and LS-THC-MP3a
essentially lie along the $x$-axis. Errors for for LS-THC-MP3a, LS-THC-MP3c,
and for standard density fitting are extrapolated to $n=40$ from
the linear region of the curves.}

\end{figure}

Density fitting and LS-THC errors for water clusters are depicted
in \figref{Errors-for-water}. Density fitting errors are exceptionally
linear, and error cancellation between the DF-MP2 and DF-MP3 correlation
energy contributions leads to a rather modest total correlation energy
error of approximately -8 micro-Hartree per correlated electron. MP2a
and MP3a errors are, for the grid size used, below 0.005 kcal/mol---an
entirely negligible amount. It is clear that rather smaller grids
could be used for these calculations (although LS-THC-MP3a is not
efficient due to its dependence on $L^{2}$) while still maintaining
manageable errors. The hand-optimized grid of Ref.~\citenum{kokkilaschumacherTensorHypercontractionSecondOrder2015}
is indeed smaller than the grid used here, but based on additional
results below, it seems likely that even smaller grids could be used
if they are appropriately pruned from a large starting grid. Since
MP2 is not a focus of this work we defer such a determination to future
work.

The LS-THC methods that include factorization of the first- and/or
second-order doubles amplitudes display larger errors, but still roughly
in line with the errors due to density fitting. These errors are noticeably
less linear with increasing system size, especially for the smaller
clusters. MP3b is the most stable, and the error is quite small, with
approximately 0.25 kcal/mol error ($7\times$ smaller than the density-fitting
error) at $(\ce{H2O})_{40}$. Approximation of the second-order amplitudes
in MP3c and MP3d, or likewise approximation of the first-order amplitudes
in MP2b incurs larger errors, but still of a manageable size. The
similarity of the MP3c and MP3d errors shows that double-fitting does
not introduce a significant additional error. Interestingly, the LS-THC-MP3
errors seems to follow separate linear trends above and below $(\ce{H2O})_{20}.$
This may be explained by the fact that larger clusters are poorly-representative
of isolated molecular clusters, including many ``broken'' hydrogen
bonds, etc. The error seems to be larger in this case, possibly indicating
that LS-THC-MP3 (with factorization of the amplitudes) is less effective
at capturing the remaining dispersion interactions compared to stronger
effects such as hydrogen bonding.

\subsection{Linear Alkanes}

\begin{figure}
\begin{tikzpicture}
\pgfplotsset{every axis plot/.append style={very thick}}
\begin{axis}[
scale=1.5,
xmin=1, xmax=40,
ymin=-2, ymax=2,
xlabel={$n$},
ylabel={Error (kcal/mol)},
legend style={at={(1.05,0.5)},anchor=west,legend cell align=right}
]
\addplot[draw=orange] table[x=n,y=DF-MP3] {alkanes-err.dat};
\addlegendentry{DF-MP3}
\addplot[draw=orange,dotted,domain=19:40,forget plot] {0.0581*x+0.0164};
\addplot[draw=blue] table[x=n,y=DF-MP2] {alkanes-err.dat};
\addlegendentry{DF-MP2}
\addplot[draw=blue,dotted,domain=19:40,forget plot] {0.0405*x-0.0007};
\addplot[draw=cyan] table[x=n,y=MP3b] {alkanes-err.dat};
\addlegendentry{LS-THC-MP3b}
\addplot[draw=yellow] table[x=n,y=MP3a] {alkanes-err.dat};
\addlegendentry{LS-THC-MP3a}
\addplot[draw=yellow,dotted,domain=20:40,forget plot] {0.0103*x-0.0264};
\addplot[draw=black!50!white] table[x=n,y=MP2a] {alkanes-err.dat};
\addlegendentry{LS-THC-MP2a}
\addplot[draw=red] table[x=n,y=MP2b] {alkanes-err.dat};
\addlegendentry{LS-THC-MP2b}
\addplot[draw=green!50!black] table[x=n,y=MP3c] {alkanes-err.dat};
\addlegendentry{LS-THC-MP3c}
\addplot[draw=green!50!black,dotted,domain=20:40,forget plot] {-0.0518*x+0.2813};
\addplot[draw=purple] table[x=n,y=MP3d] {alkanes-err.dat};
\addlegendentry{LS-THC-MP3d}
\end{axis}
\end{tikzpicture}

\caption{\label{fig:Errors-for-linear-alkanes}Errors for linear alkanes, $\text{C}_{n}\text{H}_{2n+2}$.
Note that legend entries are given in the same order as the error
values at $n=20$, and that the curve for LS-THC-MP2a essentially
lies along the $x$-axis. Errors for for LS-THC-MP3a, LS-THC-MP3c,
and for standard density fitting are extrapolated to $n=40$ from
the linear region of the curves. Data not available for $n=40$.}
\end{figure}
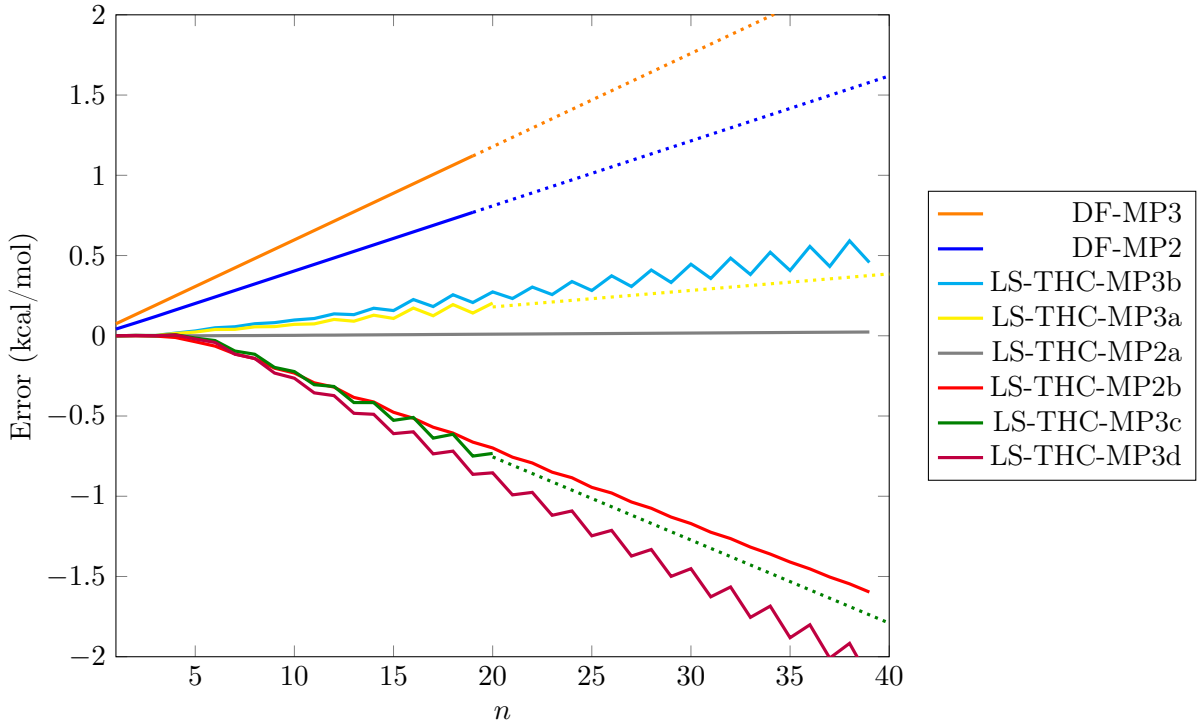
Linear alkane geometries were generated with fixed bond lengths and
angles, in order to provide a highly systematic aliphatic benchmark.
The increased linearity of the errors in \figref{Errors-for-linear-alkanes}
compared to \figref{Errors-for-water} seems to reflect this situation.
However, there are interesting deviations between the even- and odd-numbered
chains for LS-THC-MP3, and to a much more minor extent, MP2b. These
chains have point group symmetry $\text{C}_{2\text{v}}$ and $\text{C}_{2\text{h}}$,
respectively, and slightly different standard orientations. We posit
that this difference in orientation (relative to the axis-fixed molecular
grid) is responsible for the different errors, especially with respect
to the $\mathbf{X}^{(ab)}$ grid. Nonetheless, within each class the
errors are highly linear, and of very similar magnitudes as in the
case of water clusters. The biggest exception is MP3a, which now displays
a non-negligible error very similar to that of MP3b. Below, we show
that this is due to imbalance between the different grids, and is
not indicative of degraded performance of MP3a specifically. Note
that in these systems, the errors due to the DF-MP2 and DF-MP3 energy
corrections are additive, with a total correlation error of approximately
25 micro-Hartree per correlated electron. In this case, we also see
very similar errors between MP2b and MP3c/MP3d. It is tempting to
expect similar errors based on the realization that, in these methods,
we are equivalently making an LS-THC approximation of the exchange
integrals. However, results for both water clusters and linear alkenes
(next section) show no such agreement. Instead, the numerical effect
of the LS-THC approximation on the exchange integrals seems to depend
on the quantity that it is multiplied by, either the first- or second-order
amplitudes in MP2b and MP3c/MP3d, respectively. These amplitudes must
then ``sample'' different portions of the exchange integral tensor.

\subsection{Linear Alkenes}

\begin{figure}
\begin{tikzpicture}
\pgfplotsset{every axis plot/.append style={very thick}}
\begin{axis}[
scale=1.5,
xmin=1, xmax=40,
ymin=-2, ymax=2,
xlabel={$n$},
ylabel={Error (kcal/mol)},
legend style={at={(1.05,0.5)},anchor=west,legend cell align=right}
]
\addplot[draw=orange] table[x=n,y=DF-MP3] {alkenes-err.dat};
\addlegendentry{DF-MP3}
\addplot[draw=orange,dotted,domain=20:40,forget plot] {0.0486*x+0.0166};
\addplot[draw=blue] table[x=n,y=DF-MP2] {alkenes-err.dat};
\addlegendentry{DF-MP2}
\addplot[draw=blue,dotted,domain=20:40,forget plot] {0.0313*x+0.0058};
\addplot[draw=black!50!white] table[x=n,y=MP2a] {alkenes-err.dat};
\addlegendentry{LS-THC-MP2a}
\addplot[draw=cyan] table[x=n,y=MP3b] {alkenes-err.dat};
\addlegendentry{LS-THC-MP3b}
\addplot[draw=yellow] table[x=n,y=MP3a] {alkenes-err.dat};
\addlegendentry{LS-THC-MP3a}
\addplot[draw=yellow,dotted,domain=20:40,forget plot] {-0.0145*x+0.0738};
\addplot[draw=red] table[x=n,y=MP2b] {alkenes-err.dat};
\addlegendentry{LS-THC-MP2b}
\addplot[draw=green!50!black] table[x=n,y=MP3c] {alkenes-err.dat};
\addlegendentry{LS-THC-MP3c}
\addplot[draw=green!50!black,dotted,domain=20:40,forget plot] {-0.066*x+0.3323};
\addplot[draw=purple] table[x=n,y=MP3d] {alkenes-err.dat};
\addlegendentry{LS-THC-MP3d}
\end{axis}
\end{tikzpicture}

\caption{\label{fig:Errors-for-linear-alkenes}Errors for linear alkenes, $\text{C}_{n}\text{H}_{n+2}$.
Note that legend entries are given in the same order as the error
values at $n=20$, and that the curve for LS-THC-MP2a essentially
lies along the $x$-axis. Errors for for LS-THC-MP3a, LS-THC-MP3c,
and for standard density fitting are extrapolated to $n=40$ from
the linear region of the curves.}
\end{figure}
Linear (all-E) alkene geometries were also constructed with fixed
bond lengths and angles. The errors are depicted in \figref{Errors-for-linear-alkenes},
and overall show essentially the same trends as for linear alkanes,
except that now all systems are the same symmetry and so there are
no issues with orientation. The only substantive differences are that
a) the MP3a/MP3b errors are of the opposite sign (but similar magnitude),
and b) the MP2b error is somewhat smaller, and more in line proportionately
with the results for water clusters. The similar magnitude of all
LS-THC errors for alkenes compared to alkanes is encouraging, as it
shows that LS-THC-MP3 does not experience any particular issues due
to extensive electron delocalization. This highlights the advantage
of maintaining a coherent approximation of the global Hamiltonian
and/or wavefunction as opposed to approaches based on fragmentation.
For alkenes, as for the other two sets of test systems, the LS-THC
errors are seen to be highly linear, but with an interesting ``threshold''
effect. The errors for all THC-MP3 methods are essentially zero below
20--25 correlated electrons. We attribute this to crossover between
the grid size used (which scales linearly with molecular size) and
the ``exact'' grid size, that is the size of grid needed to exactly
represent the original tensors (which scales quadratically with molecular
size). The ramifications of this effect are explored in the next section.

\subsection{Size-Extensivity and Size-Consistency}

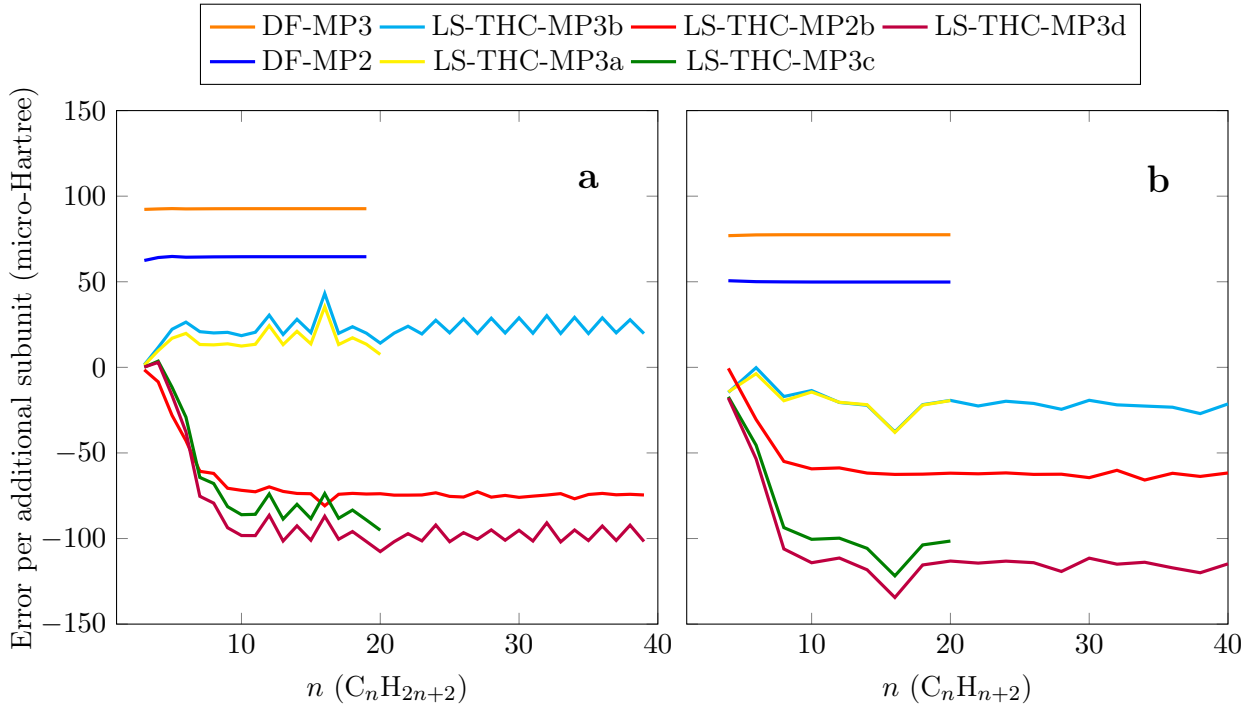
\begin{figure}
\vspace{3.5em}
\begin{tikzpicture}
\pgfplotsset{every axis plot/.append style={very thick}}
\begin{groupplot}[
y post scale=1.2,
x post scale=1.05,
xmin=1, xmax=40,
ymin=-150, ymax=150,
xlabel={$n$ ($\text{C}_{n}\text{H}_{2n+2}$)},
group style={group size=2 by 1,horizontal sep=1em},
ylabel={Error per additional subunit (micro-Hartree)}
]
\nextgroupplot
\addplot[draw=orange] table[x=n,y=DF-MP3] {alkanes-ext.dat};
\addplot[draw=blue] table[x=n,y=DF-MP2] {alkanes-ext.dat};
\addplot[draw=cyan] table[x=n,y=MP3b] {alkanes-ext.dat};
\addplot[draw=yellow] table[x=n,y=MP3a] {alkanes-ext.dat};
\addplot[draw=red] table[x=n,y=MP2b] {alkanes-ext.dat};
\addplot[draw=green!50!black] table[x=n,y=MP3c] {alkanes-ext.dat};
\addplot[draw=purple] table[x=n,y=MP3d] {alkanes-ext.dat};
\node at (axis cs:35,110) {\large{\textbf{a}}};
\nextgroupplot[
yticklabels={},
ylabel={},
xlabel={$n$ ($\text{C}_{n}\text{H}_{n+2}$)},
legend style={overlay,at={(-0.03,1.05)},anchor=south,legend cell align=right},
transpose legend=true,
legend columns=2
]
\addplot[draw=orange] table[x=n,y=DF-MP3] {alkenes-ext.dat};
\addlegendentry{DF-MP3}
\addplot[draw=blue] table[x=n,y=DF-MP2] {alkenes-ext.dat};
\addlegendentry{DF-MP2}
\addplot[draw=cyan] table[x=n,y=MP3b] {alkenes-ext.dat};
\addlegendentry{LS-THC-MP3b}
\addplot[draw=yellow] table[x=n,y=MP3a] {alkenes-ext.dat};
\addlegendentry{LS-THC-MP3a}
\addplot[draw=red] table[x=n,y=MP2b] {alkenes-ext.dat};
\addlegendentry{LS-THC-MP2b}
\addplot[draw=green!50!black] table[x=n,y=MP3c] {alkenes-ext.dat};
\addlegendentry{LS-THC-MP3c}
\addplot[draw=purple] table[x=n,y=MP3d] {alkenes-ext.dat};
\addlegendentry{LS-THC-MP3d}
\node at (axis cs:35,110) {\large{\textbf{b}}};
\end{groupplot}
\end{tikzpicture}

\caption{\label{fig:Incremental-errors-for}Incremental errors for each additional
$\ce{CH2}$ (alkanes) or $\tfrac{1}{2}\ce{C2H2}$ (alkenes) subunit.
Note that legend entries are given in the same order as the error
values at $n=20$. For both alkenes (a) and alkanes (b), the difference
between error values at $n$ and $n-2$, divided by 2, was used as
the incremental error (see text for details).}
\end{figure}
\begin{table}
\begin{tabular}{|>{\centering}p{3cm}|>{\centering}p{1cm}|>{\centering}p{1cm}|>{\raggedleft}p{1.25cm}@{\extracolsep{0pt}.}p{1.25cm}|>{\raggedleft}p{1.25cm}@{\extracolsep{0pt}.}p{1.25cm}|>{\raggedleft}p{1.25cm}@{\extracolsep{0pt}.}p{1.25cm}|>{\raggedleft}p{1.25cm}@{\extracolsep{0pt}.}p{1.25cm}|}
\hline 
Method & \multicolumn{2}{>{\centering}p{2cm}|}{Threshold (electrons)} & \multicolumn{4}{>{\centering}p{5cm}|}{Size-Consistency Error\\
(kcal/mol)} & \multicolumn{4}{>{\centering}p{5cm}|}{Incremental Error\\
(micro-Hartree/electron)}\tabularnewline
\hline 
\hline 
DF-MP2 & 0 & -1 & 0&00 & -0&01 & 10&8 & 10&0\tabularnewline
\hline 
DF-MP3 & -2 & -2 & -0&02 & -0&02 & 15&5 & 15&5\tabularnewline
\hline 
LS-THC-MP2a & 36 & 45 & 0&00 & 0&00 & 0&2 & 0&0\tabularnewline
\hline 
LS-THC-MP2b & 29 & 27 & -0&23 & -0&21 & -12&4 & -12&4\tabularnewline
\hline 
LS-THC-MP3a & 20 & 25 & 0&04 & -0&07 & 3&0 & -4&6\tabularnewline
\hline 
LS-THC-MP3b & 18 & 27 & 0&04 & -0&08 & 3&9 & -4&5\tabularnewline
\hline 
LS-THC-MP3c & 33 & 25 & -0&28 & -0&33 & -13&8 & -21&0\tabularnewline
\hline 
LS-THC-MP3d & 33 & 24 & -0&34 & -0&36 & -16&4 & -23&2\tabularnewline
\hline 
\end{tabular}

\caption{\label{tab:Size-extensivity-characteristics}Size-extensivity characteristics
as measured by linear extrapolation of alkane and alkene chains. The
first value in each column is for alkanes, while the second is for
alkenes.}
\end{table}

While canonical MP2 and MP3 are manifestly size-extensive due to the
linked structure of their diagrammatic representations, it is possible
for size-extensivity error to be introduced by further approximations.
In particular, since the LS-THC decomposition reduces the quadratic
pair molecular orbital space to a linear grid representation, we might
expect a non-linear size dependence to the incurred error and hence
a loss of size-extensivity. We have numerically measured the size-extensivity
characteristics of the methods used via the synthetic alkane and alkene
benchmarks. For each type of system, since all bond angles and distances
are fixed, the asymptotic effect of introducing a further subunit
($\ce{CH2}$ for alkenes, $\ce{C2H2}$ for alkenes) should be a constant
increase in correlation energy. Thus, if we compute the difference
between successive correlation energies as we increase the number
of subunits, we should observe an asymptotically constant value. We
further break down the successive energy differences by computing
the difference between DF and LS-THC errors of successive systems,
so that any size-extensivity error (which the canonical method does
not have) will be magnified and readily evident. Such curves are depicted
in \figref{Incremental-errors-for}. For alkanes, we have separately
computed differences between even- and odd-numbered chains and then
divided by two in order to eliminate the orientation problem. Likewise,
the alkene anergy differences have been divided by two to put them
on an equal footing (i.e. we use a $\frac{1}{2}\ce{C2H2}$ subunit).

From this figure, we do indeed see that all methods reach an essentially
constant incremental energy value. For DF-MP2 and DF-MP3, the convergence
to this value is extremely rapid and stable. For other methods, the
steady-state value is only reached for 10 or more carbon atoms. This
is a consequence of the threshold effect observed in previous sections.
Thus, both density fitting and LS-THC are size-extensive as they display
the correct asymptotic behavior. Despite the asymptotic behavior,
the threshold effect does indeed imply a small size-consistency error.
In \tabref{Size-extensivity-characteristics}, we estimate this error
as the negative of the $y$ intercept for a linear fit of the errors
in \figref{Errors-for-linear-alkanes} and \figref{Errors-for-linear-alkenes}
(only values for $n\ge8$ were used in the fit). The $x$ intercept
of this line is taken as the threshold value (as a number of correlated
electrons), while the slope gives the asymptotic incremental error
values. The $y$ intercept is essentially the anomalous energy extrapolated
to an empty system; we take its negative to indicate the expected
error in a reaction like $\text{A}+\text{B}\rightarrow\text{AB}$.
In fact, we have also numerically confirmed this error by performing
a super-system calculation of two octane molecules separated by 100
$\lyxmathsym{\AA}$. The error in comparison to twice the energy of
a single octane molecule closely matches the estimated values. However,
it is critically important to note that, unlike in truncated CI methods,
this size-consistency error is fixed and does not grow with system
size. From \tabref{Size-extensivity-characteristics}, we can also
see that the threshold effect impacts MP2a and MP3a, which do not
feature LS-THC decomposition of the amplitudes. While the size-consistency
errors are very small in these cases, they are non-zero and should
not be disregarded. For consistency we have also computed threshold
values and size-consistency errors for DF-MP2 and DF-MP3 in the same
manner; all values are essentially zero as expected.

\begin{figure}
\vspace{3.5em}
\begin{tikzpicture}
\pgfplotsset{every axis plot/.append style={very thick}}
\begin{groupplot}[
y post scale=1,
x post scale=1.05,
xmin=0, xmax=1,
ymin=0, ymax=1,
xticklabels={},
xlabel={Grid (smallest to largest)},
group style={group size=2 by 1,horizontal sep=1em},
legend style={overlay,at={(0.5,1.05)},anchor=south,legend cell align=right},
legend columns=3,
ylabel={Grid size ratio (see legend)}
]
\nextgroupplot
\addplot[draw=black] table[x expr=\thisrow{Grid}/17,y=nab] {octane-grids1.dat};
\addplot[draw=blue] table[x expr=\thisrow{Grid}/17,y=nai] {octane-grids1.dat};
\addplot[draw=green!50!black] table[x expr=\thisrow{Grid}/17,y=nij] {octane-grids1.dat};
\addplot[draw=black,dashed] table[x expr=(\thisrow{Grid}-1)/20,y=nab] {octane-grids2.dat};
\addlegendentry{$n_{ab}/n_v^2$}
\addplot[draw=blue,dashed] table[x expr=(\thisrow{Grid}-1)/20,y=nai] {octane-grids2.dat};
\addlegendentry{$n_{ai}/(n_o n_v)$}
\addplot[draw=green!50!black,dashed] table[x expr=(\thisrow{Grid}-1)/20,y=nij] {octane-grids2.dat};
\addlegendentry{$n_{ij}/n_o^2$}
\node at (axis cs:0.13,0.87) {\large{\textbf{a}}};
\nextgroupplot[
xticklabels={},
xmin=0, xmax=1,
ymin=0, ymax=10,
yticklabel pos=right,
ylabel={},
legend style={overlay,at={(0.5,1.05)},anchor=south,legend cell align=right},
legend columns=3
]
\addplot[draw=black] table[x expr=\thisrow{Grid}/17,y=nab'] {octane-grids1.dat};
\addplot[draw=blue] table[x expr=\thisrow{Grid}/17,y=nai'] {octane-grids1.dat};
\addplot[draw=green!50!black] table[x expr=\thisrow{Grid}/17,y=nij'] {octane-grids1.dat};
\addplot[draw=black,dashed] table[x expr=(\thisrow{Grid}-1)/20,y=nab'] {octane-grids2.dat};
\addlegendentry{$n_{ab}/n_{\text{DF}}$}
\addplot[draw=blue,dashed] table[x expr=(\thisrow{Grid}-1)/20,y=nai'] {octane-grids2.dat};
\addlegendentry{$n_{ai}/n_{\text{DF}}$}
\addplot[draw=green!50!black,dashed] table[x expr=(\thisrow{Grid}-1)/20,y=nij'] {octane-grids2.dat};
\addlegendentry{$n_{ij}/n_{\text{DF}}$}
\node at (axis cs:0.13,8.7) {\large{\textbf{b}}};
\end{groupplot}
\end{tikzpicture}

\caption{\label{fig:grids}Grid size ratios for each of the three pruned grids
(specific to the $ab$, $ai$, and $ij$ MO distributions), varying
the size of of the parent grid from (7,19,11) to (23,51,43) with a
fixed cutoff $\epsilon=10^{-5}$ (solid lines, the hand-optimized
grid of Ref.~\citenum{kokkilaschumacherTensorHypercontractionSecondOrder2015}
is also included), and varying $\epsilon$ logarithmically from $10^{-1}$
to $10^{-5}$ in steps of 0.2 log units with a fixed (23,51,43) parent
grid (dotted lines). In (a) the grid size ratios are given w.r.t.
the full size of the MO distributions, and in (b) the ratios are given
w.r.t. the number of density fitting functions, $n_{\text{DF}}$.}
\end{figure}
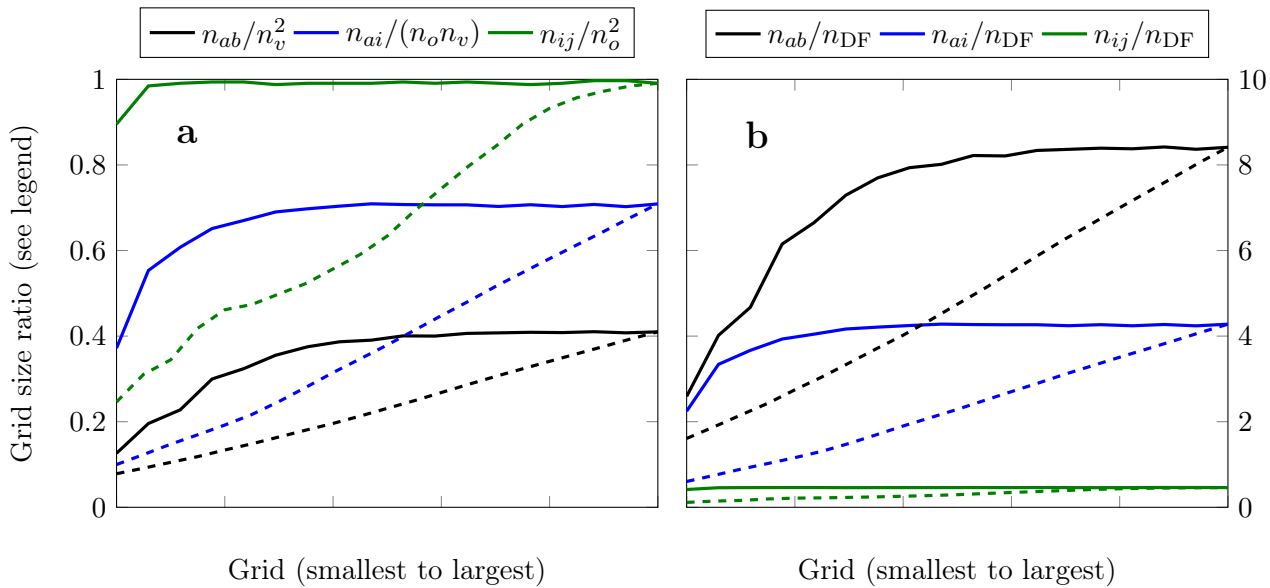
\begin{figure}[H]
\vspace{3em}
\begin{tikzpicture}
\pgfplotsset{every axis plot/.append style={
very thick,
}}
\pgfplotsset{every axis/.append style={
xmin=2.2, xmax=4.3,
ymin=-1, ymax=1,
y post scale=1.05,
x post scale=1.05,
ylabel near ticks,
scaled ticks=false,
/pgf/number format/.cd,
fixed,
/tikz/.cd
}}

\begin{axis}[
name=plot1,
ylabel={Error (kcal/mol)},
xticklabels={}
]
\addplot[draw=black,dotted] table[x=nai',y=MP3a] {octane-grids1.dat};
\addplot[draw=blue] table[x=nai',y=Ija] {octane-grids1.dat};
\addplot[draw=red,dashed] table[x=nai',y=Aia] {octane-grids1.dat};
\addplot[draw=green!50!black,dashdotted] table[x=nai',y=Aba] {octane-grids1.dat};
\node at (axis cs:4,0.7) {\large{\textbf{a}}};
\end{axis}

\begin{axis}[
name=plot2,
at={($(plot1.east)+(1em,0)$)},anchor=west,
xticklabels={},
yticklabels={}
]
\addplot[draw=black,dotted] table[x=nai',y=MP3b] {octane-grids1.dat};
\addplot[draw=blue] table[x=nai',y=Ijb] {octane-grids1.dat};
\addplot[draw=red,dashed] table[x=nai',y=Aib] {octane-grids1.dat};
\addplot[draw=green!50!black,dashdotted] table[x=nai',y=Abb] {octane-grids1.dat};
\node at (axis cs:4,0.7) {\large{\textbf{b}}};
\end{axis}

//

\begin{axis}[
name=plot3,
at={($(plot1.south)+(0,-1em)$)},anchor=north,
xlabel={$n_{ai}/n_{\text{DF}}$},
ylabel={Error (kcal/mol)}
]
\addplot[draw=black,dotted] table[x=nai',y=MP3c] {octane-grids1.dat};
\addplot[draw=blue] table[x=nai',y=Ijc] {octane-grids1.dat};
\addplot[draw=red,dashed] table[x=nai',y=Aic] {octane-grids1.dat};
\addplot[draw=green!50!black,dashdotted] table[x=nai',y=Abc] {octane-grids1.dat};
\node at (axis cs:4,0.7) {\large{\textbf{c}}};
\end{axis}

\begin{axis}[
name=plot4,
at={($(plot1.south east)+(1em,-1em)$)},anchor=north west,
xlabel={$n_{ai}/n_{\text{DF}}$},
legend style={overlay,at={($(0,2)+(-0.5em,2em)$)},anchor=south,legend cell align=right},
legend columns=4,
yticklabels={}
]
\addplot[draw=black,dotted] table[x=nai',y=MP3d] {octane-grids1.dat};
\addlegendentry{Total error}
\addplot[draw=blue] table[x=nai',y=Ijd] {octane-grids1.dat};
\addlegendentry{$\mathbf{X}^{(ij)}$ Only}
\addplot[draw=red,dashed] table[x=nai',y=Aid] {octane-grids1.dat};
\addlegendentry{$\mathbf{X}^{(ai)}$ Only}
\addplot[draw=green!50!black,dashdotted] table[x=nai',y=Abd] {octane-grids1.dat};
\addlegendentry{$\mathbf{X}^{(ab)}$}
\node at (axis cs:4,0.7) {\large{\textbf{d}}};
\end{axis}

\end{tikzpicture}

\caption{\label{fig:grids1}Errors in the total LS-THC-MP3 energy correction
and for select energy components for (a) LS-THC-MP3a, (b) LS-THC-MP3b,
(c) LS-THC-MP3c, and (d) LS-THC-MP3d, as the parent grid is varied
and with fixed $\epsilon=10^{-5}$. The energy components are labeled
by which collocation matrices appear in the expression for the middle
Hamiltonian vertex: ``$\mathbf{X}^{(ij)}$ Only'' refers to $E_{\text{HHC}}+E_{\text{HHX}}$,
``$\mathbf{X}^{(ai)}$ Only'' to $E_{\text{PH1}}+E_{\text{PH4}}+E_{\text{PH5}}$,
and ``$\mathbf{X}^{(ab)}$'' to $E_{\text{PPC}}+E_{\text{PPX}}+E_{\text{PH2}}+E_{\text{PH3}}+E_{\text{PH6}}$.
Note that in the latter case $\mathbf{X}^{(ij)}$ appears as well
in some terms, but the error due to $\mathbf{X}^{(ab)}$ is expected
to dominate. }
\end{figure}
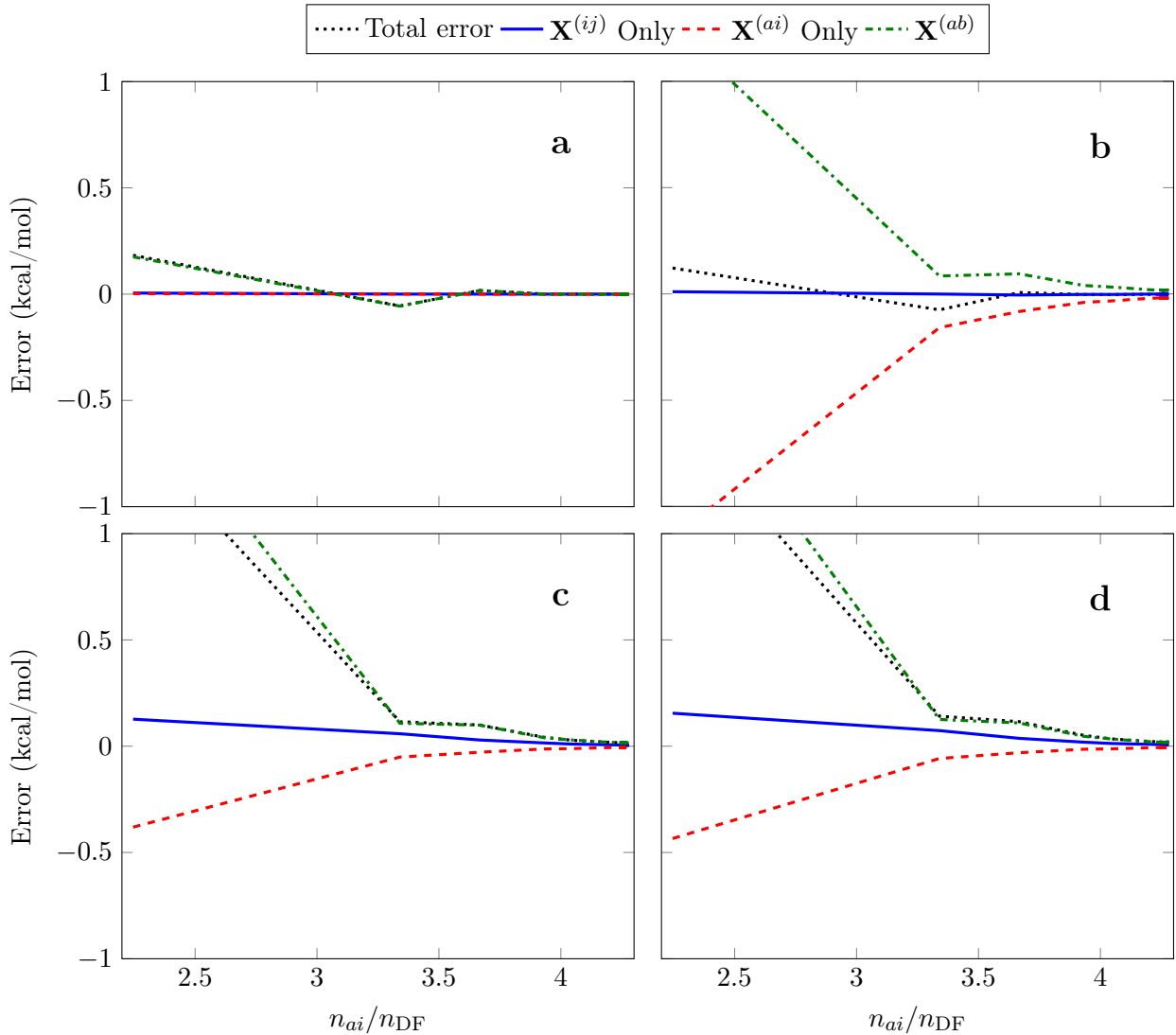
\begin{figure}[H]
\vspace{3em}
\begin{tikzpicture}
\pgfplotsset{every axis plot/.append style={
very thick,
}}
\pgfplotsset{every axis/.append style={
xmin=2.2, xmax=4.3,
ymin=-1, ymax=1,
y post scale=1.05,
x post scale=1.05,
ylabel near ticks
}}

\begin{axis}[
name=plot1,
ylabel={Error (kcal/mol)},
xticklabels={}
]
\addplot[draw=black,dotted] table[x=nai',y=MP3a] {octane-grids2.dat};
\addplot[draw=blue] table[x=nai',y=Ija] {octane-grids2.dat};
\addplot[draw=red,dashed] table[x=nai',y=Aia] {octane-grids2.dat};
\addplot[draw=green!50!black,dashdotted] table[x=nai',y=Aba] {octane-grids2.dat};
\node at (axis cs:4,0.7) {\large{\textbf{a}}};
\end{axis}

\begin{axis}[
name=plot2,
at={($(plot1.east)+(1em,0)$)},anchor=west,
xticklabels={},
yticklabels={}
]
\addplot[draw=black,dotted] table[x=nai',y=MP3b] {octane-grids2.dat};
\addplot[draw=blue] table[x=nai',y=Ijb] {octane-grids2.dat};
\addplot[draw=red,dashed] table[x=nai',y=Aib] {octane-grids2.dat};
\addplot[draw=green!50!black,dashdotted] table[x=nai',y=Abb] {octane-grids2.dat};
\node at (axis cs:4,0.7) {\large{\textbf{b}}};
\end{axis}

//

\begin{axis}[
name=plot3,
at={($(plot1.south)+(0,-1em)$)},anchor=north,
xlabel={$n_{ai}/n_{\text{DF}}$},
ylabel={Error (kcal/mol)}
]
\addplot[draw=black,dotted] table[x=nai',y=MP3c] {octane-grids2.dat};
\addplot[draw=blue] table[x=nai',y=Ijc] {octane-grids2.dat};
\addplot[draw=red,dashed] table[x=nai',y=Aic] {octane-grids2.dat};
\addplot[draw=green!50!black,dashdotted] table[x=nai',y=Abc] {octane-grids2.dat};
\node at (axis cs:4,0.7) {\large{\textbf{c}}};
\end{axis}

\begin{axis}[
name=plot4,
at={($(plot1.south east)+(1em,-1em)$)},anchor=north west,
xlabel={$n_{ai}/n_{\text{DF}}$},
legend style={overlay,at={($(0,2)+(-0.5em,2em)$)},anchor=south,legend cell align=right},
legend columns=4,
yticklabels={}
]
\addplot[draw=black,dotted] table[x=nai',y=MP3d] {octane-grids2.dat};
\addlegendentry{Total error}
\addplot[draw=blue] table[x=nai',y=Ijd] {octane-grids2.dat};
\addlegendentry{$\mathbf{X}^{(ij)}$ Only}
\addplot[draw=red,dashed] table[x=nai',y=Aid] {octane-grids2.dat};
\addlegendentry{$\mathbf{X}^{(ai)}$ Only}
\addplot[draw=green!50!black,dashdotted] table[x=nai',y=Abd] {octane-grids2.dat};
\addlegendentry{$\mathbf{X}^{(ab)}$}
\node at (axis cs:4,0.7) {\large{\textbf{d}}};
\end{axis}

\end{tikzpicture}

\caption{\label{fig:grids2}Errors in the total LS-THC-MP3 energy correction
and for select energy components for (a) LS-THC-MP3a, (b) LS-THC-MP3b,
(c) LS-THC-MP3c, and (d) LS-THC-MP3d, as $\epsilon$ is varied logarithmically
between $10^{-1}$ and $10^{-5}$ with fixed (23,51,43) parent grid.
The energy components are labeled by which collocation matrices appear
in the expression for the middle Hamiltonian vertex.}
\end{figure}
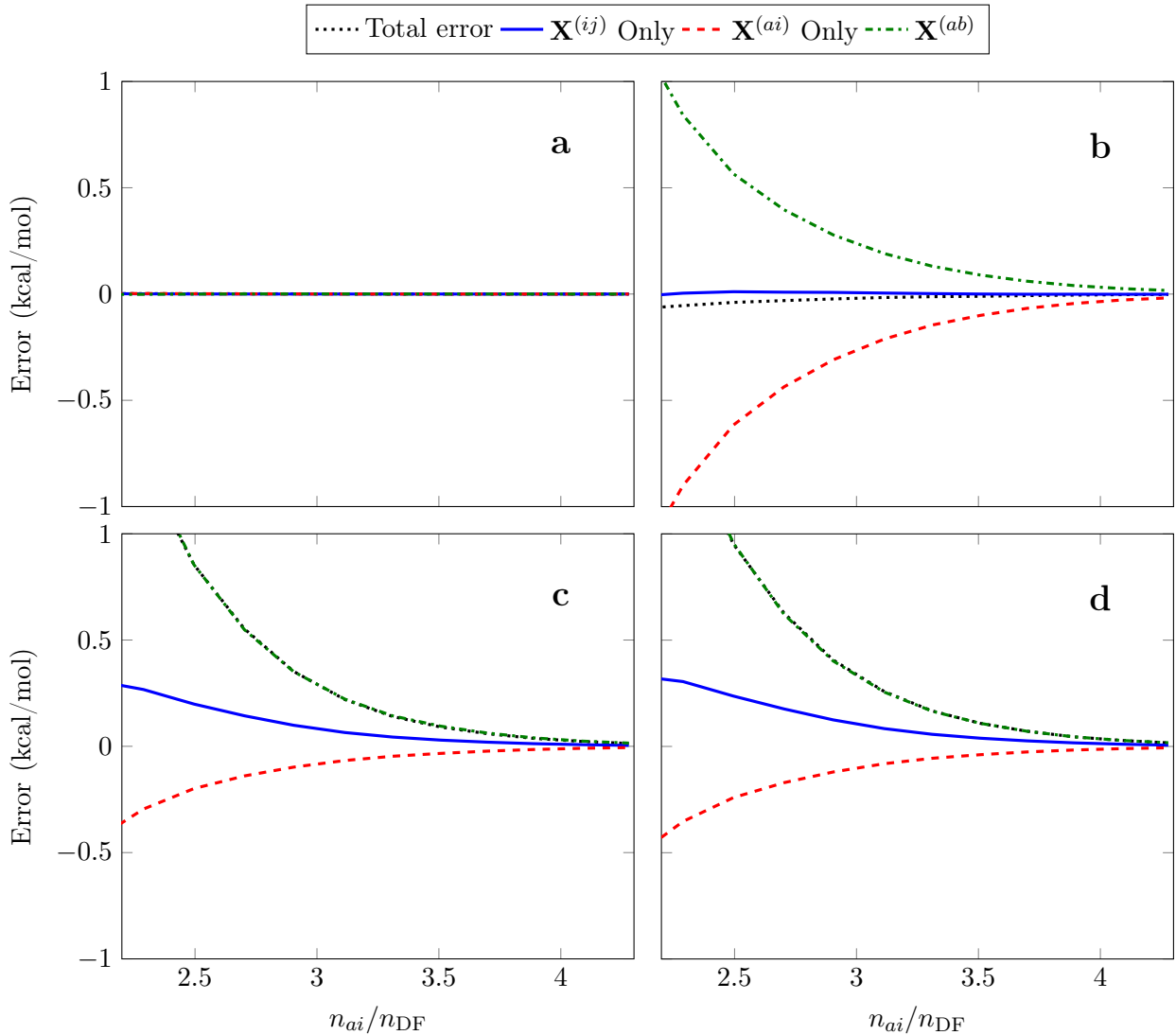
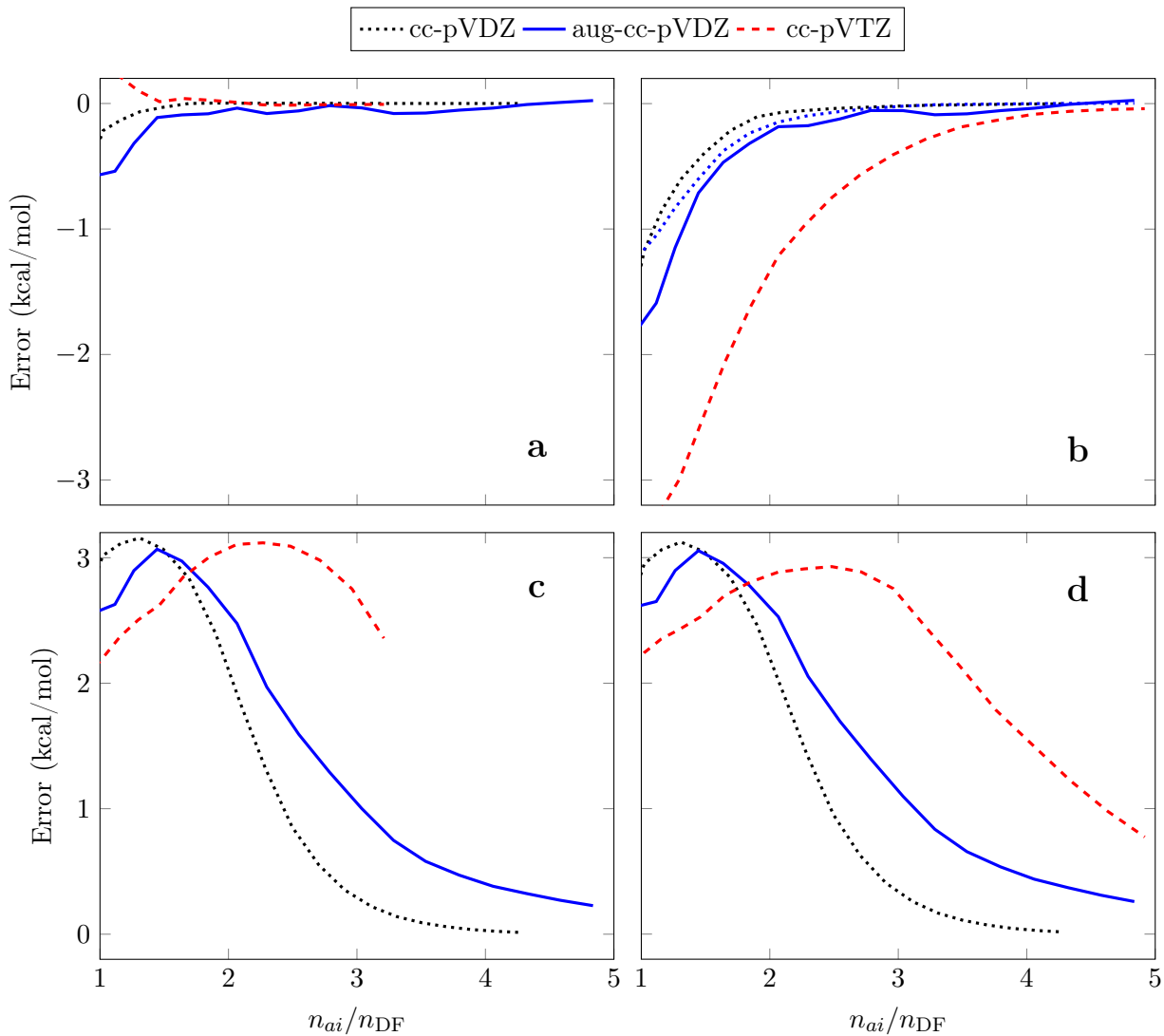
\begin{figure}[H]
\vspace{3em}
\begin{tikzpicture}
\pgfplotsset{every axis plot/.append style={
very thick,
}}
\pgfplotsset{every axis/.append style={
xmin=1, xmax=5,
ymin=-3.2, ymax=0.2,
y post scale=1.05,
x post scale=1.05,
ylabel near ticks,
scaled ticks=false,
/pgf/number format/.cd,
fixed,
/tikz/.cd
}}

\begin{axis}[
name=plot1,
ylabel={Error (kcal/mol)},
xticklabels={}
]
\addplot[draw=black,dotted] table[x=nai',y=MP3a] {octane-grids2.dat};
\addplot[draw=blue] table[x=nai',y=MP3a] {octane-grids6.dat};
\addplot[draw=red,dashed] table[x=nai',y=MP3a] {octane-grids4.dat};
\node at (axis cs:4.4,-2.75) {\large{\textbf{a}}};
\end{axis}

\begin{axis}[
name=plot2,
at={($(plot1.east)+(1em,0)$)},anchor=west,
xticklabels={},
yticklabels={}
]
\addplot[draw=black,dotted] table[x=nai',y=MP3b] {octane-grids2.dat};
\addplot[draw=blue] table[x=nai',y=MP3b] {octane-grids6.dat};
\addplot[draw=red,dashed] table[x=nai',y=MP3b] {octane-grids4.dat};
\node at (axis cs:4.4,-2.75) {\large{\textbf{b}}};
\addplot[draw=blue,dotted] table[x=nai',y expr=\thisrow{MP3b}-\thisrow{MP3a}] {octane-grids6.dat};
\end{axis}

//

\begin{axis}[
name=plot3,
ymin=-0.2, ymax=3.2,
at={($(plot1.south)+(0,-1em)$)},anchor=north,
xlabel={$n_{ai}/n_{\text{DF}}$},
ylabel={Error (kcal/mol)}
]
\addplot[draw=black,dotted] table[x=nai',y=MP3c] {octane-grids2.dat};
\addplot[draw=blue] table[x=nai',y=MP3c] {octane-grids6.dat};
\addplot[draw=red,dashed] table[x=nai',y=MP3c] {octane-grids4.dat};
\node at (axis cs:4.4,2.75) {\large{\textbf{c}}};
\end{axis}

\begin{axis}[
name=plot4,
ymin=-0.2, ymax=3.2,
at={($(plot1.south east)+(1em,-1em)$)},anchor=north west,
xlabel={$n_{ai}/n_{\text{DF}}$},
legend style={overlay,at={($(0,2)+(-0.5em,2em)$)},anchor=south,legend cell align=right},
legend columns=4,
yticklabels={}
]
\addplot[draw=black,dotted] table[x=nai',y=MP3d] {octane-grids2.dat};
\addlegendentry{cc-pVDZ}
\addplot[draw=blue] table[x=nai',y=MP3d] {octane-grids6.dat};
\addlegendentry{aug-cc-pVDZ}
\addplot[draw=red,dashed] table[x=nai',y=MP3d] {octane-grids4.dat};
\addlegendentry{cc-pVTZ}
\node at (axis cs:4.4,2.75) {\large{\textbf{d}}};
\end{axis}

\end{tikzpicture}

\caption{\label{fig:grids-tz}Basis set dependence of the total LS-THC-MP3
error using grids determined as in \figref{grids2}: (a) LS-THC-MP3a,
(b) LS-THC-MP3b, (c) LS-THC-MP3c, (d) LS-THC-MP3d. The blue dotted
line in (b) is the difference between the MP3b and MP3a energies with
aug-cc-pVDZ (see text).}
\end{figure}

\subsection{Grid and Basis Set Dependence}

The tests in the preceding section all used a fixed grid and pruning
cutoff ($\epsilon$) value. This grid was chosen to balance error
and computational efficiency (see next section), but it is instructive
to evaluate a wider range of grids, in particular for LS-THC-MP3 methods
other than MP3a. We use octane for all experiments in this section
as it is large enough to overcome the error threshold behavior, but
small enough to experiment with large grids and larger basis sets.
In \figref{grids} the essential size characteristics of two sets
of grids are depicted as applied to octane, using ``grid size ratios'',
or the number of grid points as a fraction of the total (quadratic)
pair molecular orbital space (left sub-figure) or the number of density
fitting functions (right sub-figure). The solid lines correspond to
a set of 17 regular parent grids of increasing size, from (7,19,11)
to (23,51,43), with a fixed cutoff of $\epsilon=10^{-5}$. The largest
grid is similar in size to the SG-1 grid which is commonly used in
DFT calculations. In addition, we have included Grid 0, which is the
hand-optimized grid of Ref.~\citenum{kokkilaschumacherTensorHypercontractionSecondOrder2015}.
The grid size ratios depicted in \figref{grids} are for the cc-pVDZ
basis set; we use the same grids for the cc-pVTZ studies later in
this section although of course the grid size ratios are smaller except
for the $ij$ distribution. Additionally, we have used a set of 21
grids prepared by taking the (23,51,43) parent grid and varying $\epsilon$
logarithmically between $10^{-1}$ and $10^{-5}$; the grid size ratios
are depicted by the dashed lines.

In \figref{grids1} the errors for MP3a--MP3d are depicted with respect
to the $ai$ grid size ratio. In addition to the total error in the
MP3 energy correction, we have split the error into three terms depending
on which grids are involved. In particular, we split out grids that
depend only on $\mathbf{X}^{(ij)}$ in the central Hamiltonian vertex,
as we expect this term to be very small based on the grid size ratios
in \figref{grids}. We then split the remaining terms into those which
depend on $\mathbf{X}^{(ab)}$ and those that do not (``$\mathbf{X}^{(ai)}$
only''), as the $ab$ grid is the most incomplete based on the size
ratios. The MP3a errors are all quite small, with only the $\mathbf{X}^{(ab)}$
error reaching a significant level. Individual error components (except
the $\mathbf{X}^{(ij)}$ only terms) for MP3b are significantly larger,
with errors as large or larger than 1 kcal/mol. However, these errors
efficiently cancel, leaving a total MP3b error very similar to that
of MP3a. Individual error components for MP3c and MP3d are of a similar
magnitude to MP3b, but do not cancel completely. There is some error
cancellation in this case, though, now between the $\mathbf{X}^{(ij)}$
only (no longer insignificant) and $\mathbf{X}^{(ai)}$ only terms.
This leaves the total error essentially equal to the $\mathbf{X}^{(ab)}$
error.

As the grids become larger, the MP3a errors essentially go to zero,
as was observed previously for LS-THC-MP2. The MP3b total error also
becomes quite small, although measurably non-zero. In contrast, the
individual error components, as well as total MP3c and MP3d errors
do not go to zero in the range of grid sizes tested, and remain in
the vicinity of 0.01--0.02 kcal/mol for the largest grids used. While
this error is of course quite small, it is a critical difference in
comparison to MP2a and MP3a: the LS-THC error for non-local operators
is typically larger and more slowly-decaying than for local operators.
The error cancellation properties of the various methods are readily
apparent for large grids, but are somewhat less reliable for the smaller
grids. Especially in the case of MP3b, this sharply limits the accuracy
for smaller parent grids.

Errors using the second set of grids are depicted in \figref{grids2}.
Here, the behavior for large grids is essentially the same as before,
and so is not emphasized. In contrast to the case of varying the parent
grid, keeping a large parent grid and instead varying $\epsilon$
seems to result in much smoother and more orderly error trends. Most
importantly, we can see that for grid sizes similar to the smaller
grids in \figref{grids1} (e.g. Grids 1--4), the error cancellation
properties are much better preserved. This has the effect of drastically
reducing the total error for MP3b even with similar pruned grid sizes.
For very small grids, the error cancellation properties do again break
down, but very small errors for MP3a and MP3b are still achievable.
Looking at the comparison of grid size ratios in \figref{grids},
it seems clear that the technique of controlling grid size via $\epsilon$
results in $ab$, $ai$, and $ij$ grids that are mutually balanced
and provide the greatest opportunity for effective error cancellation.
The mechanism of error cancellation in MP3b is not well understood,
except that we can note that, unlike MP3c, MP3d, as well as MP2b,
the energy expression is not written as a contraction between a set
of amplitudes and the Hamiltonian. In that case, we noted that approximating
the amplitudes is equivalent to instead approximating the exchange
integrals which might be expected to have rather severe numerical
effects. Instead, MP3b may better represent the effect of approximating
$\mathbf{T}_{2}^{[1]}$ alone, without invoking any such approximation
of the exchange integrals. While $\mathbf{T}_{2}^{[1]}$ is still
non-local (and hence leads to a non-zero MP3b error even with large
grids), the non-locality seems to be fairly limited, at least in these
examples.

Finally, we examine errors with the larger aug-cc-pVDZ and cc-pVTZ
basis sets in \figref{grids-tz}, using the second set of grids (varying
$\epsilon$). Here, apart from the small error for aug-cc-pVDZ which
we attribute to grid imbalance (most likely the parent grid requires
larger atomic radii to better represent the diffuse functions), the
MP3a error is essentially the same as for cc-pVDZ (that is to say,
quite small). For MP3b, the errors for cc-pVTZ are approximately $5\times$
larger in the small to intermediate grid size regime. In the large
grid limit, however, errors are 40--$50\times$ larger. On the other
hand, errors for aug-cc-pVDZ are essentially the same. This is especially
apparent when one considers $E_{\text{MP3b}}-E_{\text{MP3a}}$ (blue
dotted line) which cancels out the error due to grid imbalance. While
MP3b still enjoys quite effective error cancellation, using a larger
grid has a rather extreme effect on the MP3c and MP3d total energy
errors, which are $\sim45\times$ larger in the large grid limit ($\sim0.75$
kcal/mol). Decreasing $\epsilon$ only provides limited improvement
(for cc-pVTZ the error may be decreased moderately to 0.33 kcal/mol),
as values below $\sim2\times10^{-6}$ result in divergence due to
the increasingly-poor condition number of $S$. Thus, even with a
very large and hence inefficient grid, MP3c and MP3d with cc-pVTZ
are limited in accuracy to tens of micro-Hartree error per excess
correlated electron. aug-cc-pVDZ errors are intermediate, but especially
in comparison to MP3b still rather large. Of course, this error is
due to non-locality of the amplitudes (and/or the exchange integrals!).
The massive increase in the MP3b error upon going from cc-pVDZ to
cc-pVTZ, but not when going to aug-cc-pVDZ, suggests that the angular
completeness of the orbital basis set is a driving factor. Including
higher-angular momentum basis functions leads to a better description
of angular electron correlation; this is a short-range yet non-local
effect that is not well-represented by LS-THC. While longer-range
non-local effects of course still exist (in particular dispersion),
these short-range effects seem to dominate based on these experiments.
This is good news for MP3b as short-range locality should be possible
to include via a multiplicatively-separable geminal grid that still
scales linearly with molecular size. For MP3d, such a grid would perhaps
improve the errors but is unlikely to substantially eliminate them,
as the increase in error for aug-cc-pVDZ indicates longer-range non-local
errors. However, the $E_{\text{MP3d}}-E_{\text{MP3b}}$ energy difference
may provide a correction factor for similar errors in iterative LS-THC
methods.

\subsection{Timings}

\begin{figure}
\begin{tikzpicture}
\pgfplotsset{log x ticks with fixed point/.style={
xticklabel={
\pgfkeys{/pgf/fpu=true}
\pgfmathparse{exp(\tick)}
\pgfmathprintnumber[fixed relative]{\pgfmathresult}
\pgfkeys{/pgf/fpu=false}}}}
\pgfplotsset{every axis plot/.append style={very thick}}
\begin{loglogaxis}[
scale=1.5,
xmin=10, xmax=40,
ymin=100, ymax=100000,
xlabel={$n$},
ylabel={Time (s)},
xtick={10,15,...,40},
log x ticks with fixed point,
legend style={at={(1.05,0.5)},anchor=west,legend cell align=right}
]
\addplot[draw=yellow] table[x=n,y=MP3a] {timings.dat};
\addlegendentry{LS-THC-MP3a}
\addplot[draw=green!50!black] table[x=n,y=MP3c] {timings.dat};
\addlegendentry{LS-THC-MP3c}
\addplot[draw=cyan] table[x=n,y=MP3b] {timings.dat};
\addlegendentry{LS-THC-MP3b}
\addplot[draw=purple] table[x=n,y=MP3d] {timings.dat};
\addlegendentry{LS-THC-MP3d}
\addplot[draw=orange] table[x=n,y=DF-MP3] {timings.dat};
\addlegendentry{DF-MP3}
\end{loglogaxis}
\end{tikzpicture}\caption{\label{fig:Timings-for-DF-}Timings for DF- and LS-THC-MP3 methods
for water clusters, $(\ce{H2O})_{n}$ with cc-pVDZ. Both axes are
on a logarithmic scale to highlight the polynomial scaling of each
method. Legend entries are given in the order of the timings at $n=20$.}

\end{figure}
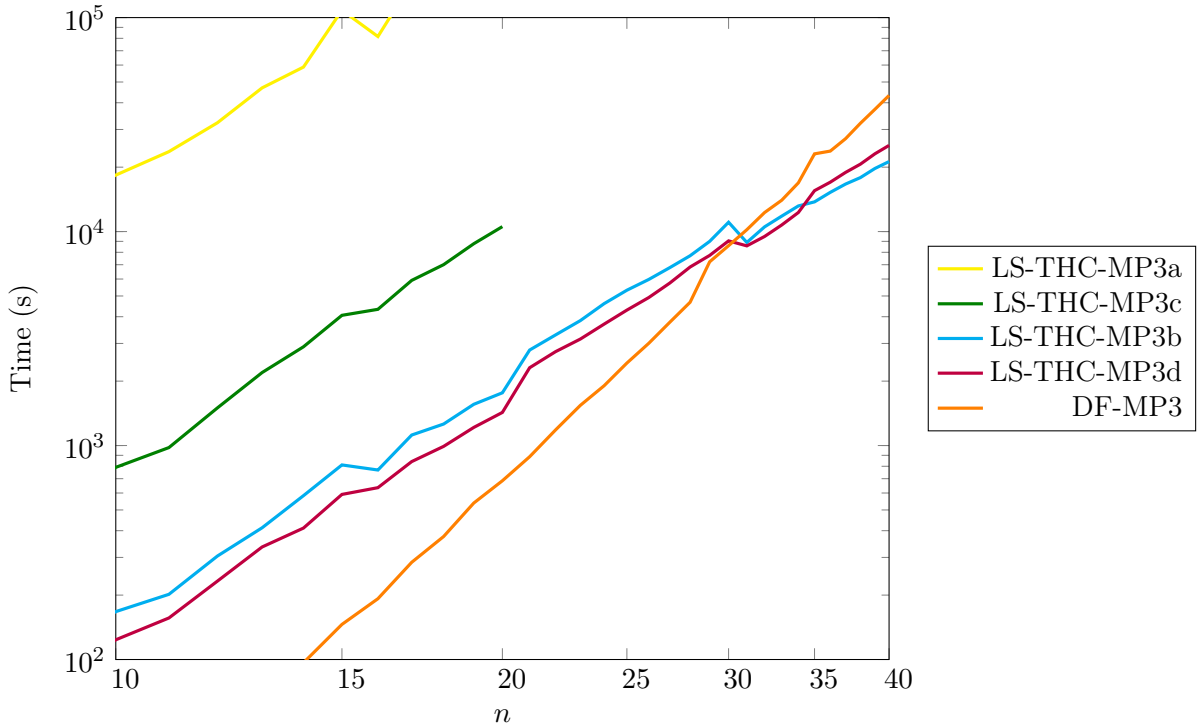

While we have shown that LS-THC-MP3 errors may be made acceptably
small while using rather compact grids, these methods are not useful
in practice unless they can also achieve a speedup over canonical
calculations. Of course, the argument with all reduced-scaling methods
is that the reduced-scaling version will ``eventually'' be faster
for large enough molecules, but how large does the molecular system
need to be? In \figref{Timings-for-DF-} we report timings for the
water cluster calculations analyzed in \subsecref{Water-Clusters}
for all LS-THC-MP3 methods, as well as for DF-MP3. The timings reported
only include the actual MP3 correlation energy calculation (and MP2,
which is a negligible component w.r.t. timings) and, for the LS-THC-MP3
calculations, the time to construct the grids and factorize the integrals.
All calculations were performed using a single node with 2x Intel
Xeon E5-2695v4 CPUs and either 256 or 768 GiB of memory; OpenMP was
used to parallelize the calculation over all 36 cores. The figure
utilizes log-log scaling of the axes, which reveals the reduced-scaling
nature of LS-THC-MP3 relative to DF-MP3. Here, the MP3b and MP3d calculations,
which have very similar computational costs, become cheaper (reach
crossover) around 240 correlated electrons. Timings for alkanes and
alkenes (not shown) exhibit a similar crossover point. While the observed
crossover point is not as small as one might hope, these results are
indeed encouraging from the standpoint of a practical LS-THC implementation,
not only of MP3 but of more elaborate methods such as CCSD as well.
The time required to construct the grids and perform the initial fit
of the integrals ranges from 3\% of the total ($n=40$) to 10\% of
the total ($n=10$). We expect further developments in reducing the
grid sizes and in improving the factorization and implementation of
LS-THC-MP3 to decrease the crossover point.

As a point of reference, the OSV-PNO-MP3 method of Hättig et al.,\citep{hattigLocalExplicitlyCorrelated2012}
which is similar to LS-THC-MP3 in that it scales as $\mathcal{O}(N^{4})$
and does not utilize orbital sparsity or domain localization, has
a crossover point with canonical MP3/cc-pVDZ at about 60 correlated
electrons. For LS-THC-MP3 to achieve the same crossover, a speedup
of approximately $10\times$ would be required. Through a combination
of improved factorization, code optimization, and reduced grid sizes,
we believe this is an eminently achievable goal; in this regard it
seems clear that both THC- and PNO-based decompositions offer a path
to efficient and effective reduced-scaling electronic structure methods,
although with very distinct theoretical and computational approaches.

\section{Conclusions}

By applying several LS-THC-MP2 and LS-THC-MP3 variants to water cluster,
linear alkane, and linear alkene tests systems, we observe a number
of trends that support the following conclusions:
\begin{enumerate}
\item With the cc-pVDZ basis set, only relatively small molecular grids
(a few hundred points per atom) are necessary to obtain LS-THC-MP3
errors similar to or smaller than errors due to density fitting. Errors
for MP2a and MP2b are essentially negligible with such grid sizes
when using balanced grids.
\item Going from cc-pVDZ to cc-pVTZ significantly increases errors for MP2b,
MP3b, MP3c, and MP3d. It does not seem to intrinsically increase the
error of MP2a and MP3a methods. Going to aug-cc-pVDZ has similar,
although less severe results, except that MP3b does not seem to be
significantly affected.
\item Starting with a large parent grid (e.g. SG-1 or even SG-0\citep{chienSG0SmallStandard2006})
and pruning the final grid size via the cutoff parameter $\epsilon$
seems to produce superior, balanced grids for a given target grid
size.
\item LS-THC methods that involve factorization of the amplitudes incur
intrinsic errors that cannot practically be eliminated by increasing
the grid size. The errors for MP3b seem to be primarily due to missing
angular correlation, which is short-range but still formally non-local.
\item The amplitude factorization in MP2b, MP3c, and MP3d is equivalent
to a factorization of the exchange integrals. This incurs a large
error as the exchange operator is highly non-local. The same is true
for any iterative LS-THC method with an ``MP2-like'' energy functional
(e.g. LS-THC-LCCD and LS-THC-CCSD).
\item LS-THC methods incur a size-consistency error due to the fact that
the error for small systems is essentially zero, but linearly increasing
above a certain threshold. This error is small, however, and does
not increase with system size.
\item Practical crossover for MP3b and MP3d is achieved around 240 correlated
electrons. Smaller crossover may be expected with further development.
\end{enumerate}
These conclusions highlight several areas for potential improvement
in LS-THC methods, but also indicate that LS-THC-MP3b, at the very
least, is promising as a practical method.
\begin{acknowledgments}
This work was supported by a generous start-up grant from SMU and
in part by the National Science Foundation (OAC 2003931). All calculations
were performed on the ManeFrame II system at the SMU Center for Scientific
Computation.
\end{acknowledgments}

\section*{Supplementary Material}

An electronic supplementary information file is available as zip file
(.zip). This file contains the the raw canonical, DF-, and LS-THC-MP$n$
correlation energies, timings, and grid parameters (in Microsoft Excel,
.xlsx), as well as molecular geometries (.xyz).

\section*{Data Availability}

The conclusions in this work are supported by data included in the
manuscript and in supporting information files available online.

\bibliographystyle{unsrt}
\bibliography{thc-mp3}

\end{document}